\documentclass[preprint,a4paper]{elsarticle}

\usepackage{graphicx}
\usepackage{amsmath}
\usepackage{float}
\usepackage{amsfonts}
\usepackage{a4wide}
\usepackage{algorithm}
\usepackage{algorithmicx}
\usepackage{algpseudocode}
\usepackage{epstopdf}
\usepackage{bm}
\usepackage{tikz}
\usepackage{multirow}
\usepackage{amsmath}
\usetikzlibrary{bayesnet}
\usetikzlibrary{shapes.gates.logic.US,trees,positioning,arrows}
\usepackage{caption}
\usepackage{subcaption}
\usetikzlibrary{trees}
\usepackage{graphicx}
\usetikzlibrary{	trees}
\usepackage{mathtools}
\usepackage{amsmath}

\usepackage{amsthm}

\usepackage{epstopdf}
\usepackage{amssymb}
\usepackage{microtype}
\usepackage{url}
\usepackage{caption}
\usepackage{subcaption}

\usepackage{xparse}

\journal{Mechanical Systems and Signal Processing}

\begin{document}
	
	\begin{frontmatter}
		
				\title{A probabilistic risk-based decision framework for structural health monitoring}
		
		\address[UoS]{Dynamics Research Group, Department of Mechanical Engineering, University of Sheffield \\ Sheffield S1 3JD, UK}
		\address[LANL]{Engineering Institute, MS T-001, Los Alamos National Laboratory, Los Alamos, NM 87545, USA}	
		
		%% Group authors per affiliation:
		\author[UoS]{A.J.\ Hughes\corref{mycorrespondingauthor}}
		\cortext[mycorrespondingauthor]{Corresponding author}
		\ead{ajhughes2@sheffield.ac.uk}
		
		\author[UoS]{R.J.\ Barthorpe}
		
		\author[UoS]{N.\ Dervilis}
		
		\author[LANL]{C.R.\ Farrar}
		
		\author[UoS]{K.\ Worden}

		\begin{abstract}
Obtaining the ability to make informed decisions regarding the operation and maintenance of structures, provides a major incentive for the implementation of structural health monitoring (SHM) systems. Probabilistic risk assessment (PRA) is an established methodology that allows engineers to make risk-informed decisions regarding the design and operation of safety-critical and high-value assets in industries such as nuclear and aerospace. The current paper aims to formulate a
risk-based decision framework for structural health monitoring that combines elements of PRA with the existing SHM paradigm. As an apt tool for reasoning and decision-making under uncertainty, probabilistic graphical models serve as the foundation of the framework. The framework involves modelling failure modes of structures as Bayesian network representations of fault trees and then assigning costs or utilities to the failure events. The fault trees allow for information to pass from probabilistic classifiers to influence diagram representations of decision processes whilst also providing nodes within the graphical model that may be queried to obtain marginal probability distributions over local damage states within a structure. Optimal courses of action for structures are selected by determining the strategies that maximise expected utility. The risk-based framework is demonstrated on a realistic truss-like structure and supported by experimental data. Finally, a discussion of the risk-based approach is made and further challenges pertaining to decision-making processes in the context of SHM are identified.
		\end{abstract}
		
		
	\end{frontmatter}
	
	%\linenumbers
	
	\section{Introduction}
	
	The field of engineering known as \textit{Structural Health Monitoring} (SHM) concerns the development and implementation of data acquisition and processing systems for the purpose of damage detection in aerospace, civil or mechanical infrastructure  \cite{Farrar2013}. A prime motivation for the use of SHM systems is to acquire the ability to make informed decisions regarding the operation and management of structures so as to improve safety and/or reduce costs. In the context of SHM, an agent tasked with making decisions is required to specify action policies that are robust to uncertainties that arise as a result of having imperfect information regarding the damage state of a structure. In addition, it may be desirable to manage a population of structures simultaneously; further complicating the decision problem. The task of decision-making for SHM is complex and highly involved and therefore demands a thorough and systematic approach.
	
	For many years, \textit{Probabilistic Risk Assessment} (PRA) has been used in industries such as aerospace and nuclear for making decisions regarding the design and operation of safety-critical assets, such as nuclear power plants \cite{USNuclearRegulatoryCommision1983} and reusable space vehicles, like the space shuttle \cite{Fragola1996}. PRA provides a rigorous and structured methodology for identifying possible adverse events associated with the operation of a system and subsequently quantifying the respective probabilities of occurrence and the severity of the consequences. 
	
	Thus far, the majority of research in the field of SHM has been focussed on the identification, localisation and classification of damage. There have been fewer attempts to address decision-making processes and to incorporate risk into SHM problems. A static probabilistic risk-based approach to SHM was applied to a simulated truss structure in \cite{Hughes2019}, although no attempt was made to forecast structural degradation. Flynn and Todd successfully applied a Bayes risk approach to the decision problem of sensor placement for an SHM system on square, gusset and T-shaped plates in \cite{Flynn2010}. The approach considered the risk of false positives and false negatives of damage identification in discrete regions of the plates. Cost-informed decision-making for miter gates was demonstrated in \cite{Vega2019}; this involved using a Bayesian neural network trained on a finite element model to infer damage and forecast performance using a transition matrix. An approach proposed in \cite{Gobbato2014}, facilitates cost-efficient reliability-based maintenance. As the latter is a reliability-based approach rather than a risk-based approach, the costs of failure events and maintenance are not  explicitly modelled. Hence, whilst the maintenance strategies developed may be cost-efficient for given safety parameters, they are not necessarily cost-optimal. There has been some research into the risk-based operation and maintenance of structures and components. Nielsen details a risk-based approach that is utilised for the operation and maintenance of off-shore wind turbines in \cite{Nielsen2013}, using probabilistic graphical modelling. Similarly, Hovgaard and Brincker provide a case study demonstrating a risk-based approach to the monitoring and maintenance of a finite element model of a wind turbine tower experiencing circumferential cracking in \cite{Hovgaard2016}. In \cite{Schobi2016} a continuous-state partially-observable Markov decision process (POMDP) was demonstrated on artificial data for maintenance planning on a deteriorating bridge. Dynamic Bayesian networks are also employed in \cite{Li2017} for the diagnosis and prognosis of the structural health of an aircraft wing; this includes the probabilistic temporal modelling and prediction of crack growth. 
	
	The current paper aims to address the lack of a generalised framework for conducting risk-based monitoring of structures at the full-system scale by augmenting the current SHM paradigm with practices employed in probabilistic risk assessment and thereby facilitating the decision-making processes that motivate the implementation of SHM systems. An overview is given of the current paradigms for conducting PRA and SHM as outlined in the literature. Also provided is background theory regarding the key technologies required for mapping PRA onto SHM; namely, probabilistic graphical models in the form of Bayesian networks and influence diagrams. A notation is established before the augmented risk-based paradigm for SHM is detailed. Finally, a discussion around the framework is made and further challenges in the SHM decision-process are identified.
	
	\section{Structural Health Monitoring}\label{sec:1}
	Structural health monitoring involves implementing damage identification strategies to determine the health state of a structure throughout its operational lifetime. Statistical pattern recognition (SPR) offers a natural approach to SHM as it allows the uncertainties inherent in engineering problems to be dealt with robustly. It is for this reason that the SPR approach has been the focus of much research over the past three decades. The established SPR paradigm for an SHM system is composed of four procedures \cite{Farrar2013}:
	
	\begin{enumerate}
		\item Operational Evaluation.
		\item Data acquisition.
		\item Feature selection.
		\item Statistical modelling for feature discrimination.
	\end{enumerate}
	
	\subsection{Operational evaluation}
	
	\textit{Operational evaluation} seeks to answer several questions concerning the implementation of an SHM system, specifically:
	
	\begin{itemize}
		\item What is the justification (safety and/or economic) for implementing an SHM system?
		\item How is damage defined for the system and what are the critical damage states?
		\item What are the environmental and operational conditions that the monitoring system is required to perform under?
		\item How does the operational environment limit data acquisition?
	\end{itemize}
	
	For an SHM system to be successfully developed and implemented, a substantial amount of information must be collected during the operational evaluation process, and significant effort may be necessary to obtain an adequate amount that is quantified in sufficient detail. Examples of required information include: monetary cost and reliability of the proposed SHM system; pertinent damage states of the structure and the thresholds at which they can deemed to have occurred, and the temperature and load variations experienced by the structure during operation.
	
	\subsection{Data acquisition}
	
	The data acquisition process is informed by the operational evaluation. The process aims to finalise the types, number and locations of sensors to be used in the SHM system. The data acquisition, storage and transmittal hardware must also be selected. The process is constrained by both economic restrictions and the limitations enforced by the expected environmental conditions; for this reason, the data acquisition process is context dependent and relies heavily on the information gathered during the operational evaluation stage.

	\subsection{Feature selection}
	
	Once the data have been acquired, a set of features must be constructed that indicate whether or not there is damage present in the structure. This procedure often involves processing the data acquired from the structure; common practices include domain transformation, dimensionality reduction, and normalisation \cite{Farrar2013}.
	
	\subsection{Statistical modelling for feature discrimination}
	
	Statistical models must be developed to exploit the discrepancy between features that indicate differing damage states. The degree of knowledge regarding the damage state obtained from an SHM system is highly dependent on the statistical model employed and can be evaluated in terms of Rytter's Hierarchy \cite{Rytter1993}:
	
	\begin{enumerate}
		\item Is there damage in the system?
		\item Where is the damage located?
		\item What type of damage is present?
		\item How severe is the damage?
		\item How much useful life remains?
	\end{enumerate}
	
	Whilst Rytter's hierarchy, in itself, does not lead to decisions being made, as SHM systems progress up the hierarchy, the information they yield becomes increasing useful to agents tasked with deciding upon a course of action for a structure. Within the field of decision theory, cost and utility are metrics used ubiquitously for the comparison of courses of action and their consequences. By combining the cost/utility of a given consequence with the respective likelihood, one can arrive at the notion of \textit{risk}.

	\section{Probabilistic Risk Assessment}\label{sec:2}
	Probabilistic risk assessment (PRA) is a method that is widely used for evaluating risks and making decisions associated with the design and management of safety-critical systems and high-value assets. In the context of PRA, risk is characterised by the likelihood of an adverse event occurring and the severity of the consequences of the event. The likelihood of occurrence for uncertain adverse events is quantified through probabilistic event sequence and system modelling. The consequences and expected costs/gains are compared and evaluated by finding an appropriate utility metric -  obvious examples include financial cost and loss of human life; however, in many applications these are overly simplistic \cite{Bedford2001}. Probabilistic risk assessment is applied in a range of industries including nuclear, aerospace and chemical process. Whilst the exact methodology used for conducting PRA differs between industries, they generally adhere to the key steps as outlined by the US Nuclear Regulatory Commission (USNRC) and the Internation Atomic Energy Agency (IAEA) \cite{USNuclearRegulatoryCommision1983}:
	
	\begin{enumerate}
		\item Initial information collection.
		\item Event-tree development.
		\item System modelling.
		\item Reliability modelling.
		\item Failure sequence quantification.
		\item Consequence analysis.
	\end{enumerate}

	Further detail is provided for each step in the following subsections.
	
	\begin{figure*}[ht!]
		\centering
		% Set the overall layout of the tree
\tikzstyle{level 1}=[level distance=2.5cm, sibling distance=2.5cm]
\tikzstyle{level 2}=[level distance=2.5cm, sibling distance=2cm]

% Define styles for bags and leafs
\tikzstyle{bag} = [text width=4em, text centered]
\tikzstyle{end} = [circle, minimum width=3pt,fill, inner sep=0pt]

% The sloped option gives rotated edge labels. Personally
% I find sloped labels a bit difficult to read. Remove the sloped options
% to get horizontal labels. 
\begin{tikzpicture}[grow=right, sloped]
\node[bag] {\textbf{Initiating event}: Jump from plane}
    child {
        node[bag] {\textbf{Top event:} Main chute}        
        child {
                node[bag] {\textbf{Top event:} Reserve chute}
                child {
                	node[end, label=right:
                	{\textbf{Adverse event:} Injury}] {}
                	edge from parent
                	node[above] {fails}
                }
                child {
                	node[end, label=right:
                	{\textbf{System successful:} Injury prevented}] {}
                	edge from parent
                	node[above] {works}
                }
                edge from parent
                node[above] {fails}
            }
            child {
                node[end, label=right:
                    {\textbf{System successful:} Injury prevented}] {}
                edge from parent
                node[above] {works}
            }
        edge from parent         
            node[above] { }
    };
\end{tikzpicture}
		\caption{An example event tree for a parachute system to prevent fall injuries \cite{USNuclearRegulatoryCommission2018}.}
		\label{fig:PRA}
	\end{figure*}
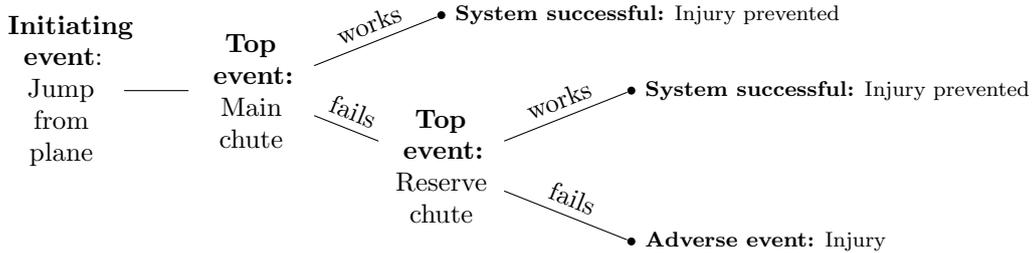
	
	\begin{figure}[ht!]
		\centering
		\includegraphics[scale=0.33]{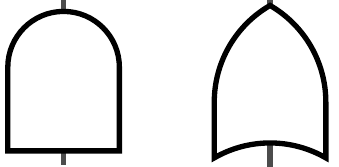}
		\caption{Fault tree diagram representations of the AND-gate (left) and OR-gate (right).}
		\label{fig:LogicGates}
	\end{figure}

	\begin{figure}[ht!]
		\centering
		\begin{tikzpicture}[
		% Gates and symbols style
		and/.style={and gate US,thick,draw,fill=white!60,rotate=90,
			anchor=east,xshift=-1mm},
		or/.style={or gate US,thick,draw,fill=white!60,rotate=90,
			anchor=east,xshift=-1mm},
		be/.style={circle,thick,draw,fill=white!60,anchor=north,
			minimum width=0.7cm},
		tr/.style={buffer gate US,thick,draw,fill=purple!60,rotate=90,
			anchor=east,minimum width=0.8cm},
		% Label style
		label distance=3mm,
		every label/.style={blue},
		% Event style
		event/.style={rectangle,thick,draw,fill=white!20,text width=2.5cm, minimum height=1.5cm,
			text centered,anchor=north},
		% Children and edges style
		edge from parent/.style={thick,draw=black!70},
		edge from parent path={(\tikzparentnode.south) -- ++(0,-1.05cm)
			-| (\tikzchildnode.north)},
		level 1/.style={sibling distance=3cm,level distance=1.4cm,
			growth parent anchor=south,nodes=event},
		level 2/.style={sibling distance=4cm},
		level 3/.style={sibling distance=3cm},
		level 4/.style={sibling distance=1cm}
		%%  For compatability with PGF CVS add the absolute option:
		%   absolute
		]
		%% Draw events and edges
		\node (g1) [event] {\textbf{Top event:} Reserve chute fails}
		child {node (g2) {\textbf{Intermediate event:} Chute not deployed}
			child {node (b8) {\textbf{Basic event:} Ripcord breaks}}
			child {node (g4) {\textbf{Intermediate event:} Auto activation fails}
				child {node (b5) {\textbf{Basic event:} Altimeter malfunctions}}
				child {node (b7) {\textbf{Basic event:} Dead battery}}
			}
		}
		child {node (b0) {\textbf{Basic event:} Chute tangled}};
		%% Place gates and other symbols
		%% In the CVS version of PGF labels are placed differently than in PGF 2.0
		%% To render them correctly replace '-20' with 'right' and add the 'absolute'
		%% option to the tikzpicture environment. The absolute option makes the
		%% node labels ignore the rotation of the parent node.
		\node [or]	at (g1.south)	[label=-20: ]	{};
		\node [and]	at (g2.south)	[label=-20: ]	{};
		\node [or]	at (g4.south)	[label=-20: ]	{};
		\node [be]  at (b0.south)	[label=below: ]	{};
		\node [be]	at (b5.south)	[label=below: ]	{};
		\node [be]	at (b7.south)	[label=below: ]	{};
		\node [be]	at (b8.south)	[label=below: ]	{};
		\end{tikzpicture}
		\caption{An example fault tree for a reserve parachute system \cite{USNuclearRegulatoryCommission2018}.}
		\label{fig:ChuteFT}
	\end{figure}
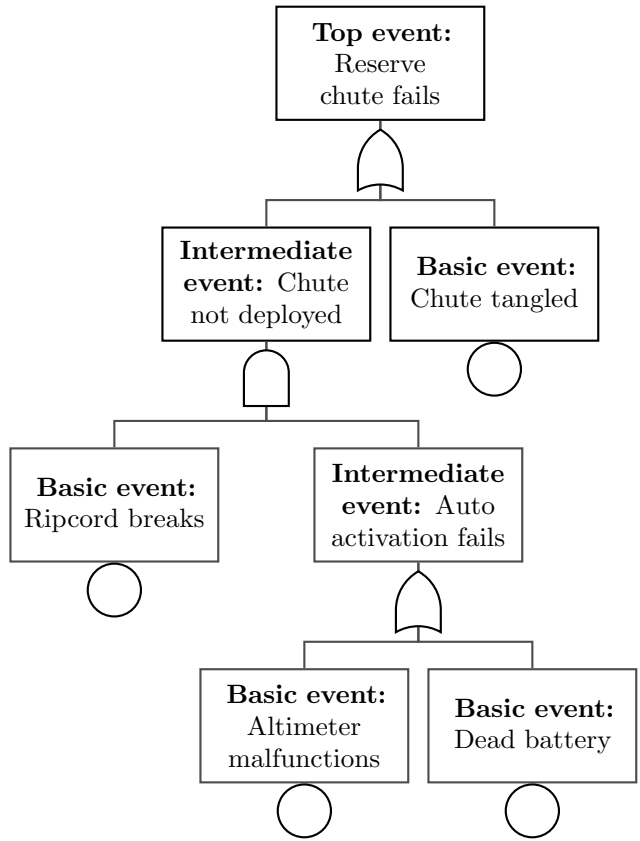
	
	\subsection{Initial information collection}
	
	Information regarding the design and operation of the structure in question is collated. Details such as component specifications, loading and environmental conditions are considered. The information gathered at this stage is used to inform the subsequent steps. Given the large quantity of information required for conducting PRA, an important factor to be considered at this stage is the method by which the necessary information is represented, stored and managed. A common practice is to utilise a database \cite{Bedford2001}. There is a clear analogy here with the operational evaluation stage for SHM.
	
	\subsection{Event-tree development}
	
	Event trees outline potential accident sequences - combinations of initiating events and the subsequent system failures or successes that may result in an adverse consequence. The sequences of system failures and successes are known as \textit{top events}. The system failures identified in the event-tree development stage are subsequently modelled as fault trees. An example event tree for a system designed to prevent injury following a jump from a plane is shown in Figure \ref{fig:PRA}.
	
	\subsection{Fault tree development}
	
	Fault trees are used in PRA to facilitate the quantification of system failure probabilities. The development of fault trees involves expressing the causal relationships between component failures and subsystem failures using Boolean logic gates. The level of detail captured in the fault tree (the level of components which are incorporated) is determined by the component level for which meaningful reliability data can be obtained. Failures of components belonging to the most fundamental level incorporated in the fault tree are known as \textit{basic events} and are represented in a fault tree diagram as circles. Intermediate and top events are defined as combinations of other intermediate, and basic events through Boolean logic gates such as the AND-gate and OR-gate. The fault tree diagram notation for the AND-gate and OR-gate are shown in Figure \ref{fig:LogicGates}. An example fault tree for the deployment of a reserve parachute is shown in Figure \ref{fig:ChuteFT}.
	
	\subsection{Reliability modelling}
	
	Information regarding the reliability of system components and the frequency of initiating events is necessary to quantify the probabilities of top events; this is typically gleaned from data and by applying appropriate reliability models.
	
	\subsection{Failure sequence quantification}
	
	By assigned the components in the fault trees with failure rates, the probability of top events may be computed. Propagating the probabilities of the initiating events and top events through the event tree allows the probability of each possible outcome in the event tree to be calculated.
	
	\subsection{Consequence analysis}
	
	With the probability of an adverse event quantified, a cost/utility metric must be chosen so that the risk associated with the failure sequence may be assessed. The risk assessment may then be used to inform design decisions, such as increasing safety by introducing additional redundancies in the system, or optimising cost by removing components that do not cause the risk to fall below an acceptable threshold. The risk assessment may also be used to inform risk-based inspection for a system that is in operation.
	
	\section{Probabilistic Graphical Models} \label{sec:3}
	Probabilistic graphical models (PGMs) are a powerful framework for reasoning and decision-making under uncertainty - a core problem in SHM. Probabilistic graphical models  are representations of joint probability distributions, in which nodes denote a set of random variables and edges connecting nodes imply dependency between variables. The probabilistic graphical model representation provides benefits over a flat (non-graphical) representation \cite{Sucar2015}:
	
	\begin{itemize}
		\item They provide a compact and intuitive representation of complex probability distributions which makes them easier to understand, communicate and learn.
		\item They facilitate efficient computation by exploiting local independence structures.
	\end{itemize}
	
	A PGM over a set of $N$ variables $X$ may be specified by a set of $M$ local functions $f(Y_i)$, where $Y_i$ is some subset of $X$, and a graph $G$ comprised of nodes/vertices $V$ and edges $E$. The joint probability distribution represented by the graph is obtained by:
	
	\begin{equation}\label{PGM}
	P(X_1, X_2, \ldots, X_N) = K \prod_{i=1}^{M}f(Y_i)
	\end{equation}
	
	\noindent where $K$ is a normalisation factor ensuring the probabilities sum to unity.
	
	There are two classes of problem associated with PGMs: \textit{inference} and \textit{learning}. Inference is concerned with obtaining the marginal or conditional probabilities of a subset of variables $Z$ given any other subset  $Y$. i.e. $P(Z|Y)$. Learning is concerned with obtaining the graph structure and parameters given a complete, or incomplete, set of observed data values for $X$. i.e.\ $G,f(Y_i)|X$. The remainder of the current paper will be primarily concerned with inference problems and their application in a risk-informed SHM framework.
	
	\subsection{Bayesian networks}
	
	Bayesian networks (BNs) are a form of PGM. Specifically, they are directed acyclic graphs (DAGs) in which nodes represent random variables and edges connecting nodes represent conditional dependencies between variables. For discrete random variables, the local functions that describe the conditional probability distributions (CPDs) between variables are conditional probability tables (CPTs), and in the case of continuous random variables are conditional probability density functions (CPDFs).
	
	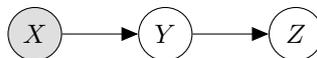
\begin{figure}[ht!]
		\centering
		\begin{tikzpicture}[x=1.7cm,y=1.8cm]
		
		% Nodes for plate GM

		\node[obs] (parent) {$X$} ;
		\node[latent, right=1cm of parent] (inter) {$Y$} ;
		\node[latent, right=1cm of inter] (child) {$Z$} ;
		
		% Edges for plate GM {from} {to}
		\edge {parent} {inter} ; %
		\edge {inter} {child} ; %
		
		\end{tikzpicture}
		\caption{An example Bayesian network.}
		\label{fig:PGM0}
	\end{figure}
	
	Figure \ref{fig:PGM0} shows a simple Bayesian network where $X$ is a parent of $Y$ and an ancestor of $Z$; $Z$ is said to be the child of $Y$ and a descendant of $X$. Node $X$ is independent of of other nodes and so is specified by the unconditional distribution $P(X)$. Observed variables are shaded grey.
	
	Given observations on a subset of nodes in a BN, inference algorithms can be applied to obtain posterior distributions over the unobserved random variables. In some cases, analytical solutions of posterior distributions may be found by using exact inference methods. To solve the inference problem using direct computation from the joint probability distribution, the computational complexity increases exponentially with the size of the graph and quickly becomes intractable. Fortunately, algorithms have been developed that allow efficient computation \cite{Pearl1986}.
	
	\subsection{Influence diagrams}
	
\begin{figure}[ht!]
		\centering
		\begin{tikzpicture}[x=1.7cm,y=1.8cm]
		
		% Nodes for plate GM

		\node[latent] (condition) {$W_c$} ;
		\node[obs,left=1cm of condition] (forecast) {$W_f$} ;
		\node[rectangle,draw=black,minimum width=0.7cm,minimum height=0.7cm,below=1cm of forecast] (decision) {$D$} ;
		\node[det, below=1cm of condition] (utility) {$U$} ;
		
		% Edges for plate GM
		\edge {condition} {forecast} ; %
		\edge {forecast} {decision} ; %
		\edge {condition} {utility} ; %
		\edge {decision} {utility} ; %
		
		\end{tikzpicture}
		\caption{An example influence diagram representing the decision of whether to go outside or stay in under uncertainty in the future weather condition given an observed forecast.}
		\label{fig:ID0}
	\end{figure}
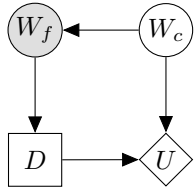

	Bayesian networks can be augmented to represent decision processes by incorporating nodes for decision variables and utility functions - these augmented networks are known as \textit{influence diagrams} \cite{Kjaerulff2008}. Decision nodes are denoted by squares and utility nodes are denoted by rhombi, as shown in Figure \ref{fig:ID0}. Edges connecting random variables to utility nodes denote that the utility function is dependent on the state of that variable. Similarly, edges connecting decision nodes to utility nodes denote that the utility function has a dependence on the action decided upon. Edges connecting random variables or decision nodes to other decision nodes denote order, that is to say the random variable or decision is observed prior to the decision being made; such edges are referred to as \textit{informational links} as they do not imply a functional dependence but rather that the information regarding the state of the variable is required for the decision to be made.
	
	The influence diagram shown in Figure \ref{fig:ID0} can be interpreted as a binary decision process regarding whether to go out for a walk or stay in and watch TV under uncertainty in the future weather condition $W_c$ given an observed weather forecast $W_f$. The nodes $W_f$ and $W_c$ can be considered as binary random variables representing the weather forecast and actual weather condition, respectively, with possible states $domain(W_f) = domain(W_c) = \big\{ bad, good \big\}$ and the weather forecast is dependent on the weather condition. The possible actions can be summarised as $domain(D) = \big\{ TV, walk \big\}$. The utility $U$ achieved is then dependent on both the weather condition experienced and the decided action.
	
	In general, a policy $\delta$ is a mapping from all possible observations to possible actions. The problem of inference in influence diagrams is to determine an optimal strategy $\mathbf{\Delta}^{\ast} = \big\{\delta_{1}^{\ast},\ldots, \delta_{n}^{\ast} \big\}$ given a set of observations on random variables where $\delta_{i}^{\ast}$ is the policy for the $i^{th}$ decision to be made in a strategy $\mathbf{\Delta}^{\ast}$ that yields the \textit{maximum expected utility} ($MEU$). The expected utility is a function of probability and utility; and by this definition is equivalent to risk.

	\section{Definitions}
	To establish a framework for mapping PRA onto SHM, some fundamental concepts will first be defined. In addition, a notation will be established for describing structures that can be expressed as hierarchical graphs.
	
	One must start with a physical structure or system of interest $\bm{S}$. It is assumed that $\bm{S}$ may be defined in terms of constituent units; components $c$, joints $j$ and substructures $\bm{s}$. Components and joints are considered irreducible base units of $\bm{S}$ whereas substructures are compound units and may be comprised of joints, components and other substructures.
	
	\begin{figure*}[ht!]
		\centering
		\begin{tikzpicture}[x=1.7cm,y=1.8cm]

		\node (S11) {$\bm{S}^{1}$} ;
		\node[below=1cm of S11] (j12) {$j_{1}^{2}$} ;
		\node[left=1cm of j12] (s12) {$\bm{s}_{1}^{2}$} ;
		\node[right=1cm of j12] (s22) {$\bm{s}_{2}^{2}$} ;
		\node[below=1cm of j12] (c13) {$c_{1}^{3}$} ;
		\node[below=1cm of s12] (j13) {$j_{1}^{3}$} ;
		\node[below=1cm of s22] (j23) {$j_{2}^{3}$} ;
		\node[left=1cm of j13] (s13) {$\bm{s}_{1}^{3}$} ;
		\node[right=1cm of j23] (s23) {$\bm{s}_{2}^{3}$} ;
		\node[below=1cm of s13] (j1L) {$j_{1}^{L}$} ;
		\node[below=1cm of s23] (j2L) {$j_{N_{j}^{L}}^{L}$} ;
		\node[left=1cm of j1L] (c1L) {$c_{1}^{L}$} ;
		\node[right=1cm of j1L] (c2L) {$c_{2}^{L}$} ;
		\node[left=1cm of j2L] (c3L) {$c_{N_{c}^{L}-1}^{L}$} ;
		\node[right=1cm of j2L] (c4L) {$c_{N_{c}^{L}}^{L}$} ;

		\draw [<-] (S11) -- (j12);
		\draw [<-] (S11) -- (s12);
		\draw [<-] (S11) -- (s22);
		\draw [<-] (s12) -- (s13);
		\draw [<-] (s12) -- (j13);
		\draw [<-] (s12) -- (c13);
		\draw [<-] (s22) -- (c13);
		\draw [<-] (s22) -- (j23);
		\draw [<-] (s22) -- (s23);
		\draw [<-, loosely dotted] (s13) -- (c1L);
		\draw [<-, loosely dotted] (s13) -- (j1L);
		\draw [<-, loosely dotted] (s13) -- (c2L);
		\draw [<-, loosely dotted] (s23) -- (c3L);
		\draw [<-, loosely dotted] (s23) -- (j2L);
		\draw [<-, loosely dotted] (s23) -- (c4L);

		\end{tikzpicture}
		\caption{A hierarchical graphical representation of a generic structure $\bm{S}$. Superscripts denote the level in the hierarchy and subscript indexes each type of constituent unit in a given level. Dotted edges imply an arbitrary structuring between levels.}
		\label{fig:Struct0}
	\end{figure*}
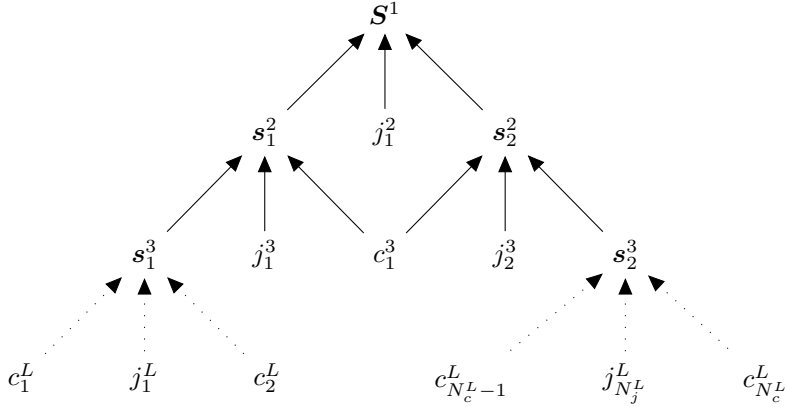
	
	Figure \ref{fig:Struct0} depicts a graphical representation of a hierarchical structure that may be considered without loss of generality. Nodes represent the global structure and its constituent units and edges represent the dependence of a (sub)structure on its constituent units. At the top, or level 1, of the hierarchy is the global structure with the hierarchy level denoted in the superscript. It can be seen that the global structure $\bm{S}^{1}$ is comprised of two substructures $\bm{s}_{1}^{2}$, $\bm{s}_{2}^{2}$ and a joint $j_{1}^{2}$, i.e.\ $\bm{S}^1 = \big\{ \bm{s}_{1}^{2},j_{1}^{2},\bm{s}_{2}^{2}\big\}$. These units form the second level of the hierarchy. $\bm{s}_{1}^{2}$ and $\bm{s}_{2}^{2}$ may in turn be expanded to yield $\bm{S}^1 = \big\{ \{\bm{s}_{1}^{3},j_{1}^{3},c_{1}^{3}\},j_{1}^{2},\{c_{1}^{3},j_{2}^{3},\bm{s}_{2}^{3}\}\big\}$. Progressing down the hierarchy levels, one can continue to expand the substructures into constituent units until the $L^{th}$ level of the hierarchy which is comprised solely of base units. By taking the expansion of $\bm{S}^{1}$ into its constituent base units and discarding the repeated units arising from substructures that share components, one obtains a list of the base units that form a given structure $\bm{S}$. Within each level of the hierarchy, units are numbered via a subscript from 1 to $N_{u}^{i}$ where $N_{u}^{i}$ is the number of a constituent unit type $u$ in the $i^{th}$ level of the hierarchy. The notation $u^{i}_{n}$ where $i$ is an integer from 1 to $L$  and $n$ is an integer from 1 to $N_{u}^{i}$ provides a unique identifier for each unit within a structure.

	It is assumed that there exists a set of features $\bm{\nu}$, observable from $\bm{S}$, that are produced according to a generative latent state model with latent state $\bm{H}^{\ast}(t)$, where $\bm{H}^{\ast}(t)$ is the true health state of $\bm{S}$ and may be expressed in terms of the true health states of the constituent components and joints  $h^{\ast}c^i_n(t)$ and $h^{\ast}j^i_n(t)$, respectively, i.e.\ $\bm{H}^{\ast}(t) = \big\{ h^{\ast}{c^{2}_1}(t),\ldots,h^{\ast}{c^{L}_{N^{L}_{c}}}(t),h^{\ast}{j^2_1}(t), \ldots,h^{\ast}{j^L_{N^L_j}}(t)\big\}$.
	
	The structure $\bm{S}$ also has a predicted time-dependent health state vector $\bm{H}(t) = \big\{ h{c^{2}_1}(t),$
	$\ldots,h{c^{L}_{N^{L}_{c}}}(t),h{j^2_1}(t), \ldots,$ $h^{j^L_{N^L_j}}(t)\big\}$. Health-state vectors can be constructed from any subset of components and joints.
	
	For the structure/system $\bm{S}$, there must exist a set of failure modes of interest $\bm{F} = \big\{ F_1,\ldots,F_{N_{F}}\big\}$ whereby $\bm{S}$ ceases to be fit for purpose. It is assumed that a given failure mode is dependent on the health states of a subset of components, joints and substructures for which a health-state vector can be constructed. In addition, each failure state has an associated utility $U_{F_{n}}$.
	
	Finally, for the structure $\bm{S}$, there also exists a set of decisions $\bm{d} = \big\{ d_1,\ldots,d_{N_{d}}\big\}$ which affect $\bm{H}^\ast$, each having an associated utility $U_{d_i}$. In addition, there will exist some set of environmental conditions $\bm{e} = \big\{ e_1(t),\ldots, e_{n_{e}}(t)\big\}$ that will alter the distribution of $\bm{\nu}$.
	
	\section{Mapping PRA onto SHM}
	Upon examination, it becomes apparent that there are both differences and similarities between the paradigms for SHM and PRA that can be examined to determine which aspects of PRA will be useful for SHM. Whilst it is clear that both SHM and PRA are utilised for the purpose of making decisions in the face of uncertainty, PRA is conducted offline for a system experiencing a set of anticipated initiating events. In contrast, the decision processes for which SHM is implemented are online and require continual predictions of the damage state of the structure. It is for this reason that the event-tree development stages and failure sequence quantification stages in PRA are less applicable to SHM.
	
	Both paradigms begin with collating information regarding the structure and defining the context in which decisions are to be made. In fact, the first three stages of the PRA paradigm involve expressing the structure and context in a logical way which facilitates the quantification of risk and the decision-making process. It is in this formal expression of the structure that the decision-making process in the SHM paradigm stands to benefit. An overview of the risk-based SHM paradigm is as follows:
	
	\begin{enumerate}
		\item Operational evaluation.
		\item Failure-mode modelling.
		\item Decision modelling.
		\item Data acquisition.
		\item Feature selection.
		\item Statistical modelling for feature discrimination.
	\end{enumerate}
	
	\subsection{Operational evaluation}
	
	With the aim of justifying the use and defining the context of a risk-based SHM system, the operational evaluation stage seeks to answer many of the same questions as in the standard paradigm. However, some questions require an approach that facilitates the failure-mode modelling and decision-modelling stages.
	
	Foremost, information regarding the components $\bm{c}$, joints $\bm{j}$, substructures $\bm{s}$, and the dependencies between them is required.
	
	When identifying the critical damage states of the structure $\bm{S}$, one should aim to identify the failure modes of interest $\bm{F}$. Critical components, joints and substructures/subsystems that contribute to $\bm{F}$ should also be identified at this stage. The predicted damage states of these components $h$ should be defined. The damage states of the critical substructures/subsystems $\bm{H}$ should be defined as a vector in terms of $h$.
	
	For each failure mode in $\bm{F}$, potential decisions $\bm{d}$ should be identified and the ways in which the actions influence the structure or likelihood of failure modes occurring should be determined. Utility values $\bm{U}_F$ and  $\bm{U}_d$ for all $\bm{F}$ and all $\bm{d}$, respectively, should be quantified. The selection of utility values will determine the behaviour of the decision-making agent, and is analogous to setting a decision threshold in a standard SHM paradigm.
	
	Environmental influences $\bm{e}$ should also be identified. It should also be decided whether the SHM system is to evaluate the health of the structure at static, independent instances in time, or predict future health states, thereby requiring a model forecasting the degradation of the structure.
	
	For large, complex structures it may be beneficial to borrow the data management techniques used in PRA, such as databases, to organise the information obtained during the Operational Evaluation stage. This will allow for a rigorous and structured approach to the information collection and allow for the identification of aspects of the SHM system that require further specification or more information. Having a formal information structure will also expedite the subsequent failure-mode modelling step which requires detailed knowledge of the physical structure.
	
	\begin{figure*}[ht!]
		\centering
		\begin{tikzpicture}[
		% Gates and symbols style
		and/.style={and gate US,thick,draw,fill=white!60,rotate=90,
			anchor=east,xshift=-1mm},
		or/.style={or gate US,thick,draw,fill=white!60,rotate=90,
			anchor=east,xshift=-1mm},
		be/.style={circle,thick,draw,fill=white!60,anchor=north,
			minimum width=0.7cm},
		tr/.style={buffer gate US,thick,draw,fill=purple!60,rotate=90,
			anchor=east,minimum width=0.8cm},
		% Label style
		label distance=3mm,
		every label/.style={blue},
		% Event style
		event/.style={rectangle,thick,draw,fill=white!20,text width=1cm,
			text centered,font=\sffamily,anchor=north},
		% Children and edges style
		edge from parent/.style={thick,draw=black!70},
		edge from parent path={(\tikzparentnode.south) -- ++(0,-1.05cm)
			-| (\tikzchildnode.north)},
		level 1/.style={sibling distance=3cm,level distance=1.4cm,
			growth parent anchor=south,nodes=event},
		level 2/.style={sibling distance=4cm},
		level 3/.style={sibling distance=3cm},
		level 4/.style={sibling distance=1cm}
		%%  For compatability with PGF CVS add the absolute option:
		%   absolute
		]
		%% Draw events and edges
		\node (g1) [event] {$F^1_1$}
		child {node (b1) {$c^2_1$}}
		child {node (g2) {$s^2_1$}
			child {node (g3) {$s^3_1$}
				child {node (b2) {$c^4_1$}}
				child {node (b3) {$j^4_1$}}
				child {node (b4) {$c^4_2$}}
			}
			child {node (b8) {$j^3_1$}}
			child {node (g4) {$s^3_2$}
				child {node (b5) {$c^4_3$}}
				child {node (b6) {$j^4_2$}}
				child {node (b7) {$c^4_4$}}
			}
		}
		child {node (b0) {$j^2_1$}};
		%% Place gates and other symbols
		%% In the CVS version of PGF labels are placed differently than in PGF 2.0
		%% To render them correctly replace '-20' with 'right' and add the 'absolute'
		%% option to the tikzpicture environment. The absolute option makes the
		%% node labels ignore the rotation of the parent node.
		\node [or]	at (g1.south)	[label=-20: ]	{};
		\node [and]	at (g2.south)	[label=-20: ]	{};
		\node [or]	at (g3.south)	[label=-20: ]	{};
		\node [or]	at (g4.south)	[label=-20: ]	{};
		\node [be]  at (b0.south)	[label=below: ]	{};
		\node [be]	at (b1.south)	[label=below: ]	{};
		\node [be]	at (b2.south)	[label=below: ]	{};
		\node [be]	at (b3.south)	[label=below: ]	{};
		\node [be]	at (b4.south)	[label=below: ]	{};
		\node [be]	at (b5.south)	[label=below: ]	{};
		\node [be]	at (b6.south)	[label=below: ]	{};
		\node [be]	at (b7.south)	[label=below: ]	{};
		\node [be]	at (b8.south)	[label=below: ]	{};
		\end{tikzpicture}
		\caption{A fault tree of a (single) failure mode $F_1$ where the superscript denotes the hierarchy level and the subscript is an identifier.}
		\label{fig:FT1}
	\end{figure*}
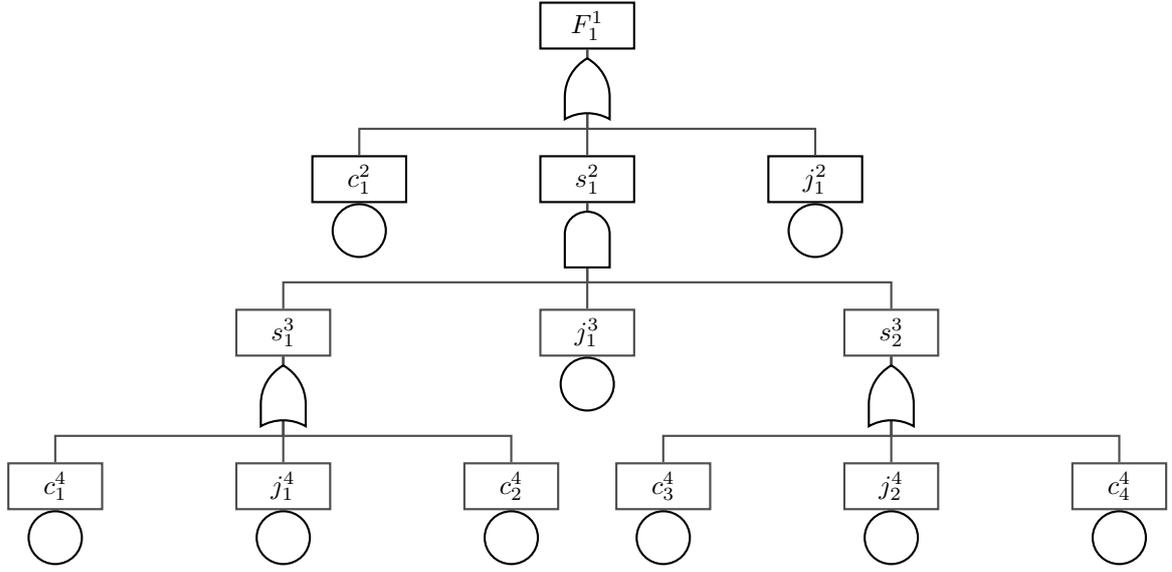
	
	\begin{figure}[ht!]
		\centering
		\begin{tikzpicture}[x=1.7cm,y=1.8cm]
		
		% Nodes for plate GM
		
		\node[latent] (failure) {$F^1_1$} ;
		\node[latent, below=1cm of failure] (sub11) {$h{s^2_1}$} ;
		\node[latent, left=1cm of sub11] (comp11) {$h{c^2_1}$} ;
		\node[latent, right=1cm of sub11] (join11) {$h{j^2_1}$} ;
		\node[latent, below=1cm of sub11] (join21) {$h{j^3_1}$} ;
		\node[latent, left=1.75cm of join21] (sub21) {$h{s^3_1}$} ;
		\node[latent, right=1.75cm of join21] (sub22) {$h{s^3_2}$} ;
		\node[latent, below=1cm of sub21] (join31) {$h{j^3_1}$} ;
		\node[latent, below=1cm of sub22] (join32) {$h{j^3_2}$} ;
		\node[latent, right=0.5cm of join31] (comp32) {$h{c^4_2}$} ;
		\node[latent, left=0.5cm of join32] (comp33) {$h{c^4_3}$} ;
		\node[latent, left=0.5cm of join31] (comp31) {$h{c^4_1}$} ;
		\node[latent, right=0.5cm of join32] (comp34) {$h{c^4_4}$} ;
		\node[latent, below=2.5cm of join21] (H) {$\bm{H}$} ;
		
		% Edges for plate GM
		
		\edge {sub11} {failure} ; %
		\edge {comp11} {failure} ; %
		\edge {join11} {failure} ; %
		\edge {sub21} {sub11} ; %
		\edge {sub22} {sub11} ; %
		\edge {comp31} {sub21} ; %
		\edge {join31} {sub21} ; %
		\edge {comp32} {sub21} ; %
		\edge {comp33} {sub22} ; %
		\edge {join32} {sub22} ; %
		\edge {comp34} {sub22} ; %
		\edge {join21} {sub11} ;
		\edge {H} {comp31} ;
		\edge {H} {comp32} ;
		\edge {H} {comp33} ;
		\edge {H} {comp34} ;
		\edge {H} {join31} ;
		\edge {H} {join32} ;
		\edge {H} {join21} ;
		\edge {H} {comp11} ;
		\edge {H} {join11} ;

		\end{tikzpicture}
		\caption{A Bayesian network of failure mode in $F_1$.}
		\label{fig:FTBN}
	\end{figure}
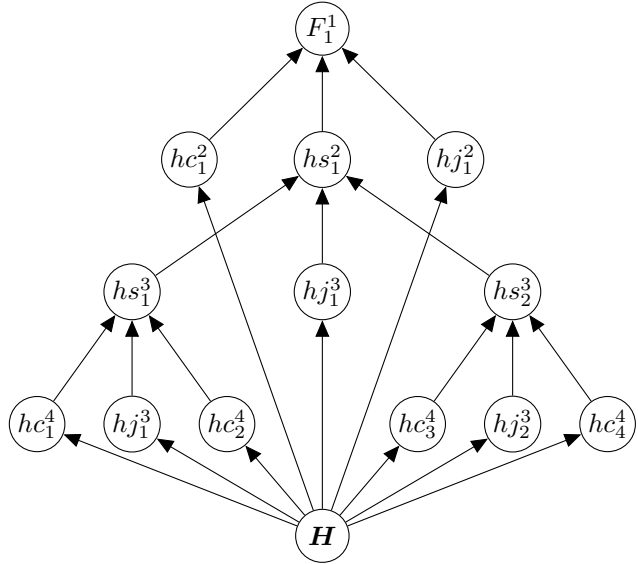
	
	\subsection{Failure-mode modelling}
	
	For each of the failure modes of interest in $\bm{F}$, one should proceed to construct a fault tree, such as that shown in Figure \ref{fig:FT1}, based upon the health states of the relevant components, joints and substructures/subsystems. It should be noted here that, in many cases, the exact nature of the failure modes will be unknown and so a best estimate based on engineering judgement may be used.
	
	Fault trees offer a rigorous and consistent structure for expressing the failure modes; however, as statements in Boolean logic they are limited in their flexibility. In the context of SHM, it is desirable to represent the components in a fault tree as having multiple damage states, and it is for this reason that one should map the constructed fault trees into Bayesian networks. Bobbio \textit{et al} outline a convenient mapping from fault trees into Bayesian networks in \cite{Bobbio2001}, whilst also highlighting the additional flexibility that is granted by doing so. Additionally, Bayesian networks are used to represent structural failures in \cite{Mahadevan2001}.
	
	In the example shown in Figure \ref{fig:FTBN}, the component health states, substructure health states and failure event are represented as random variables where the substructure health states are conditioned on the component health states and the failure events is conditioned on the substructure health state. The random variables are defined using a conditional probability distribution (CPD) which may be discrete or continuous.
	
	A node representing the health-state vector of the critical components and joints $\bm{H}$ should be included in the fault tree, as this latent state will be predicted during the statistical modelling process. To define the vector $\bm{H}$ within the Bayesian network, the conditional dependence between the nodes representing the local health states of the components and joints and  $\bm{H}$ are expressed as a binary logic table.
	
	One function of the failure-mode Bayesian network is to allow the flow of information from the statistical model to the decision, whilst parsing the information in a way that facilitates the defining of the failure events $\bm{F}$. The network also allows the computation of marginal distributions for the probability of failure in each component, joint, or substructure allowing for damage localisation.

	\subsection{Decision modelling}
	
	Modelling the decision process involves augmenting the Bayesian network developed in the failure modelling stage with nodes for each decision in $\bm{d}$ and for utilities $\bm{U}_F$ and $\bm{U}_d$ to produce an influence diagram. Decision nodes in which the actions alter the probability of a state or event should modify the CPDs accordingly. Utility nodes are constrained to be leaf nodes and should be dependent on the appropriate failure events or decisions.
	
	For static problems, it may be convenient to model the decision process in a separate influence diagram which receives information regarding failure probabilities from the fault tree. This issue is because it is implicit that the decision is made after observations are made; if one attempts to solve a network in which a decision is made that yields a state that is inconsistent with the observed state, a conflict arises.
	
	\subsection{Data acquisition}
	
	The data acquisition process should not differ from that in the standard SHM paradigm. Here, there is a subtlety that the data acquisition system should be designed so as to optimise the decision-making rather than damage identification.
	
	\subsection{Feature selection}
	
	The feature selection process should not differ from that in the standard SHM paradigm. Again, there is the subtlety that the features should be selected so as to optimise  the decision-making.
	
	\subsection{Statistical modelling}
	
	The purpose of the statistical model is to predict the critical health states $\bm{H}$ given the selected feature set $\bm{\nu}$. As aforementioned, it is assumed that $\bm{\nu}$ is produced through a generative latent state model, with latent-state $\bm{H}$. Probabilistic classifiers that output a probability distribution over all possible states of $\bm{H}$, such as Gaussian mixture models (GMMs), are compatible. Probabilistic classifiers are instrumental in building robustness to the uncertainty surrounding the true health state of $\bm{S}$ into the decision process. Ideally, the chosen statistical model will be capable of consistently identifying the actual health state under all identified operating and environmental conditions $\bm{e}$, or at least appropriately reflect the uncertainty caused by varying conditions in the prediction.
	
	Finally, if a model describing the degradation of $\bm{S}$ (i.e. a transition model for $\bm{H}$) is required for forecasting failure events in the time-dependent case, the CPDs defining $P(\bm{H}_{t}|\bm{H}_{t-1},\bm{d})$ should be specified accordingly.
	
	\section{Case study: Four-bay truss}
\begin{figure*}[h!]
		\centering
		\includegraphics*[scale=0.36]{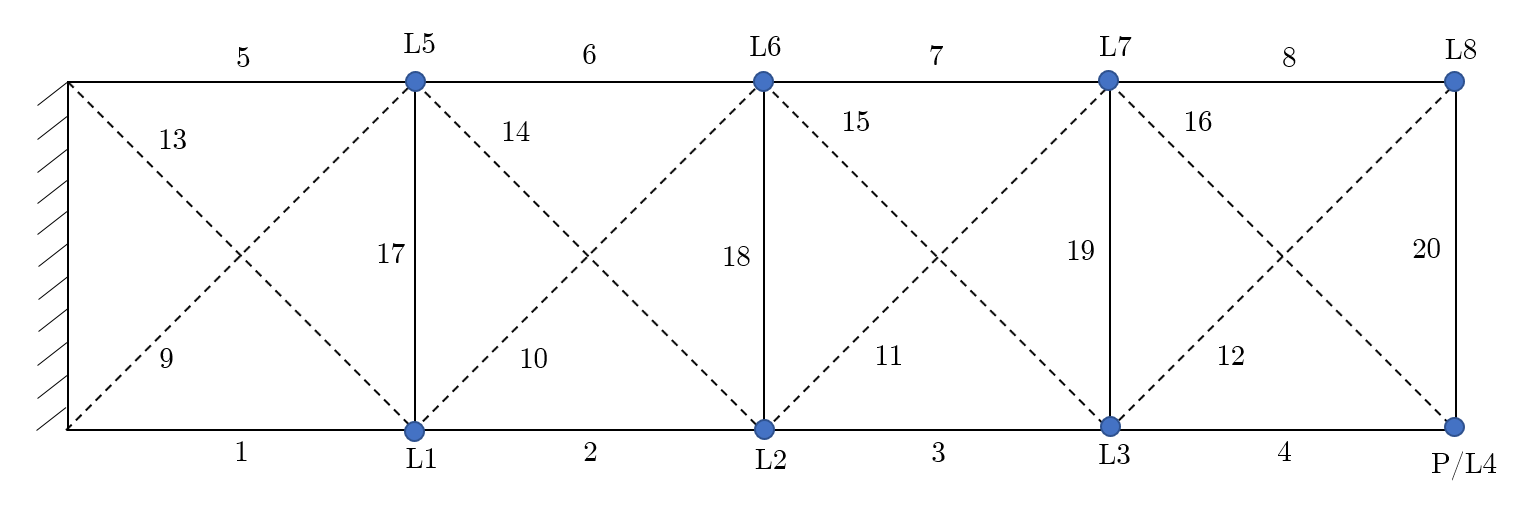}
		\caption{A 2-dimensional four-bay truss comprised of 20 members, 8 of which are removable and denoted by a dashed line. Loads are applied at points L, and a preload is applied at point P. Load positions are shown as blue dots. The bays are numbered left to right from 1 to 4.}
		\label{fig:Truss}
	\end{figure*}

	To demonstrate the probabilistic graphical model formulation of a risk-based approach to an SHM problem, it was applied to a four-bay truss identical to that used in \cite{Worden1993} and shown in Figure \ref{fig:Truss}. For clarity, the example will be limited to a single failure mode and a single binary decision. The failure of joints will also be ignored.
	
	The truss was composed of 20 aluminium members each with length 250mm and cross-sectional area 177 mm$^2$; the overall length of the structure being 1 m and the height being 0.25 m. The members were pinned together with steel bolts in lubricated holes. The truss was subjected to a preload of 5 kg at point P and additional consecutive loads of 10, 20 and 30 kg at each of the 8 points L in turn. For each of the 24 load cases, microstrains were measured at the midpoints of the 12 horizontal and vertical members. This process was performed 8 times in total, once with each cross-member removed.
	
	In addition to the experimental data, a finite element simulation of the truss was developed whereby removal of a cross-member was achieved by assigning a Young's modulus of 1 MPa. As well as the 10, 20 and 30 kg loads used in the experiment, the truss was simulated with loads of 5, 15 and 25 kg. Furthermore, strains were obtained for the truss under each load case in its undamaged condition, i.e. with all cross-members intact.
	
	\subsection{`Operational Evaluation' of Truss}
	In order to construct a risk-based decision framework for the truss, one must first define it formally. As it was elected to ignore joints, the global truss structure $\bm{T}$ can be defined as four substructures, one for each bay i.e. $\bm{T} = \big\{ \bm{b}_{1},\bm{b}_{2},\bm{b}_{3},\bm{b}_{4}\big\}$. As only the failure of cross-members was considered, each bay can in turn be defined as two components, for example, $\bm{b}_{1} = \big\{ m_{9}, m_{13} \big\}$. Consequently, the global structure may be represented as $\bm{T} = \big\{ m_{9},m_{10},m_{11},m_{12},m_{13},m_{14},m_{15},m_{16}\big\}$.
	
	A single failure mode $F_{\bm{T}}$ of the truss was considered; the full or partial collapse of the structure. This failure mode corresponds to the event where the truss is no longer able to support the load/preload, hence, $F_{\bm{T}}$ occurs when both cross-members in a single bay fail.
	
	In an attempt to minimise the occurence of the failure mode of interest $F_{\bm{T}}$, a single binary decision $d$ was identified; a choice between `do nothing' and `perform maintenance'. In addition, utilities were assigned to the failure event and the decidable actions in a manner which may reflect the relative costs associated with failure and maintenance in real-world engineering applications. The utilities assigned to the failure and decision are shown in Tables \ref{tab:UF} and \ref{tab:Ud}, respectively.
	
	\begin{table}[ht]
		\centering
		% table caption is above the table
		\caption{A table showing the entries of the utility function $U(F_{\bm{T}})$ where $F_{\bm{T}}=0$ and $F_{\bm{T}}=1$ denote the truss being operational and failed, respectively.}
		\label{tab:UF}       % Give a unique label
		% For LaTeX tables use
		\begin{tabular}{c|c}
			\hline\noalign{\smallskip}
			$F_{\bm{T}}$ & $U(F_{\bm{T}})$ \\
			\noalign{\smallskip}\hline\noalign{\smallskip}
			$0$ & $15$ \\
			$1$ & $-285$ \\
			\noalign{\smallskip}\hline
		\end{tabular}
	\end{table}
	
	\begin{table}[ht]
		\centering
		% table caption is above the table
		\caption{A table showing the entries of the utility function $U(d)$ where $d=0$ and $d=1$ denote the `do nothing' and `perform maintenance' actions, respectively.}
		\label{tab:Ud}       % Give a unique label
		% For LaTeX tables use
		\begin{tabular}{c|c}
			\hline\noalign{\smallskip}
			$d$ & $U(d)$ \\
			\noalign{\smallskip}\hline\noalign{\smallskip}
			$0$ & $0$ \\
			$1$ & $-100$ \\
			\noalign{\smallskip}\hline
		\end{tabular}
	\end{table}
	
	For the purposes of demonstration it is assumed that the load on the truss will be uncertain, varying in discrete time within the interval $[0,w_{max}]$ where $w_{max}$ is defined in subsection \ref{sec:HsTM}. Furthermore, it is assumed that the location of the load also varies in discrete time and that, in the limit of infinite time-slices, each of the 8 locations is visited an equal number of times.
	
	\subsection{Truss Failure Modelling}
	
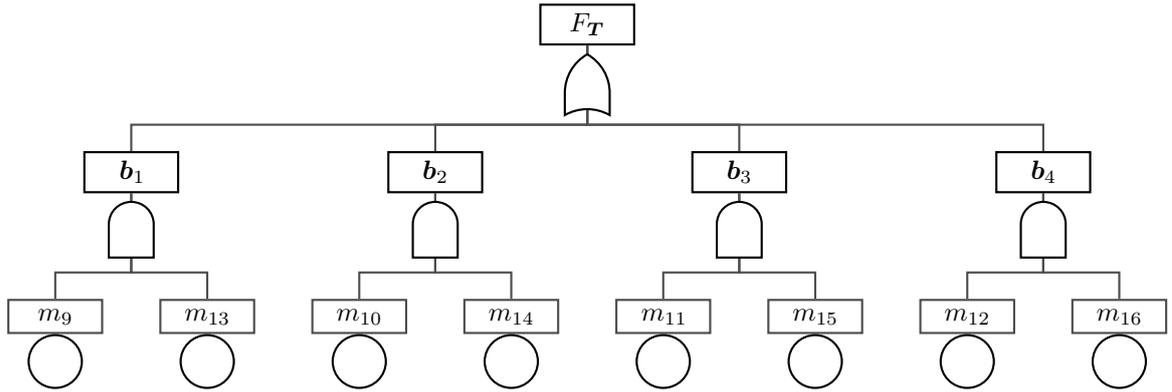
\begin{figure*}[h!]
		\centering
		\begin{tikzpicture}[scale=1,
		% Gates and symbols style
		and/.style={and gate US,thick,draw,fill=white!60,rotate=90,
			anchor=east,xshift=-1mm},
		or/.style={or gate US,thick,draw,fill=white!60,rotate=90,
			anchor=east,xshift=-1mm},
		be/.style={circle,thick,draw,fill=white!60,anchor=north,
			minimum width=0.7cm},
		% Label style
		label distance=3mm,
		every label/.style={blue},
		% Event style
		event/.style={rectangle,thick,draw,fill=white!20,text width=1cm,
			text centered,font=\sffamily,anchor=north},
		% Children and edges style
		edge from parent/.style={thick,draw=black!70},
		edge from parent path={(\tikzparentnode.south) -- ++(0,-1.05cm)
			-| (\tikzchildnode.north)},
		level 1/.style={sibling distance=4cm,level distance=1.4cm,
			growth parent anchor=south,nodes=event},
		level 2/.style={sibling distance=2cm},
		level 3/.style={sibling distance=1cm},
		level 4/.style={sibling distance=1cm}
		%%  For compatability with PGF CVS add the absolute option:
		%   absolute
		]
		%% Draw events and edges
		\node (g1) [event] {$F_{\bm{T}}$}
		child {node (g2) {$\bm{b}_1$}
			child {node (b1) {$m_{9}$}}
			child {node (b2) {$m_{13}$}}
		}
		child {node (g3) {$\bm{b}_2$}
			child {node (b3) {$m_{10}$}}
			child {node (b4) {$m_{14}$}}
		}
		child {node (g4) {$\bm{b}_3$}
			child {node (b5) {$m_{11}$}}
			child {node (b6) {$m_{15}$}}
		}
		child {node (g5) {$\bm{b}_4$}
			child {node (b7) {$m_{12}$}}
			child {node (b8) {$m_{16}$}}
		};
		%% Place gates and other symbols
		%% In the CVS version of PGF labels are placed differently than in PGF 2.0
		%% To render them correctly replace '-20' with 'right' and add the 'absolute'
		%% option to the tikzpicture environment. The absolute option makes the 
		%% node labels ignore the rotation of the parent node. 
		\node [or]	at (g1.south)	[label=-20: ]	{};
		\node [and]	at (g2.south)	[label=-20: ]	{};
		\node [and]	at (g3.south)	[label=-20: ]	{};
		\node [and]	at (g4.south)	[label=-20: ]	{};
		\node [and]	at (g5.south)	[label=-20: ]	{};
		\node [be]	at (b1.south)	[label=below: ]	{};
		\node [be]	at (b2.south)	[label=below: ]	{};
		\node [be]	at (b3.south)	[label=below: ]	{};
		\node [be]	at (b4.south)	[label=below: ]	{};
		\node [be]	at (b5.south)	[label=below: ]	{};
		\node [be]	at (b6.south)	[label=below: ]	{};
		\node [be]	at (b7.south)	[label=below: ]	{};
		\node [be]	at (b8.south)	[label=below: ]	{};
		\end{tikzpicture}
		\caption{A fault tree of failure mode $F_{\bm{T}}$ for a four-bay truss. The failure mode $F_{\bm{T}}$ occurs if at least one bay $\bm{b}$ fails. A bay $\bm{b}$ will fail if both cross-members $m$ fail.}
		\label{fig:TrussFailFT}
	\end{figure*}

	The failure mode $F_{\bm{T}}$ of the truss can be represented as the fault tree shown in Figure \ref{fig:TrussFailFT} where the failure of a bay $\bm{b}$ is defined as the and-gate of two cross-member $m$ failures, and the failure of the truss $F_{\bm{T}}$ is defined as the or-gate of the bay failures.

\begin{figure*}[h!]
		\centering
		\begin{tikzpicture}[x=1.7cm,y=1.8cm]
		
		% Nodes for plate GM
		
		\node[latent] (failure) {$F_{\bm{T}}$} ;
		\node[const, below=1cm of failure] (temp) {$ $} ;
		\node[latent, left=0.1cm of temp] (bay2) {$h\bm{b}_{2}$} ;
		\node[latent, left=1cm of temp] (bay1) {$h\bm{b}_{1}$} ;
		\node[latent, right=0.1cm of temp] (bay3) {$h\bm{b}_{3}$};
		\node[latent, right=1cm of temp] (bay4) {$h\bm{b}_{4}$} ;
		\node[const, below=1.5cm of temp] (temp2) {$ $} ;
		\node[latent, left=0.1cm of temp2] (mem6) {$hm_{14}$} ;
		\node[latent, left=1.5cm of temp2] (mem2) {$hm_{10}$} ;
		\node[latent, left=3cm of temp2] (mem5) {$hm_{13}$} ;
		\node[latent, left=4.5cm of temp2] (mem1) {$hm_{9}$} ;
		\node[latent, right=0.1cm of temp2] (mem3) {$hm_{11}$} ;
		\node[latent, right=1.5cm of temp2] (mem7) {$hm_{15}$} ;
		\node[latent, right=3cm of temp2] (mem4) {$hm_{12}$} ;
		\node[latent, right=4.5cm of temp2] (mem8) {$hm_{16}$} ;
		
		% Edges for plate GM
		\edge {bay1} {failure} ; %
		\edge {bay2} {failure} ; %
		\edge {bay3} {failure} ; %
		\edge {bay4} {failure} ; %
		\edge {mem1} {bay1} ; %
		\edge {mem5} {bay1} ; %
		\edge {mem2} {bay2} ; %
		\edge {mem6} {bay2} ; %
		\edge {mem3} {bay3} ; %
		\edge {mem7} {bay3} ; %
		\edge {mem4} {bay4} ; %
		\edge {mem8} {bay4} ; %

		\end{tikzpicture}
		\caption{A Bayesian network representation of failure mode $F_{\bm{T}}$ for a four-bay truss.}
		\label{fig:TrussFailBN}
	\end{figure*}
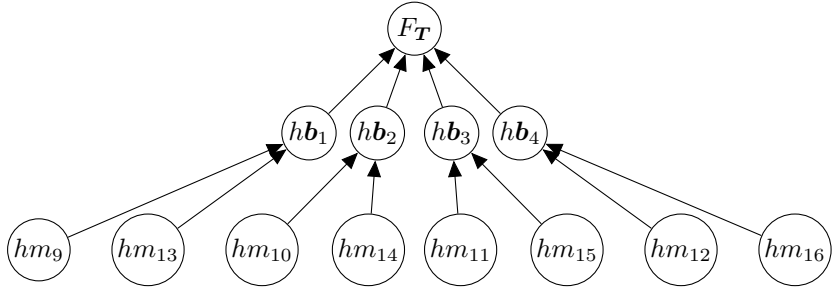

	To map the fault tree for the failure event $F_{\bm{T}}$ into a probabilistic graphical model, $h\bm{b}$ and $hm$ will be used to denote the random variables that represent the local binary health states of the bays and cross-members, respectively, where 0 corresponds to intact and 1 corresponds to failed. Additionally, $\bm{H}$ will be used to denote the random variable vector for the health state of the global structure where $\bm{H} = \big\{ hm_{9},hm_{10},hm_{11},hm_{12}, hm_{13}, hm_{14},hm_{15},hm_{16}\big\}$. For conciseness, the vector $\bm{H}$ will, on occasion, be summarised as $\bm{H} = H$, where $H$ is the decimal representation of the 8-bit binary number specified by the vector. Finally, the $F_{\bm{T}}$ notation will be retained to represent the random variable corresponding to the failure event. The Bayesian network corresponding to the fault tree shown in Figure \ref{fig:TrussFailFT} is shown in Figure \ref{fig:TrussFailBN}. The conditional probability distributions specifying $P(F_{\bm{T}}|h\bm{b}_{1},h\bm{b}_{2},h\bm{b}_{3},h\bm{b}_{4})$ (or $P(F_{\bm{T}}|\bm{hb})$ for brevity) and $P(h\bm{b}_{1}|hm_{9},hm_{13})$ (or $P(h\bm{b}_{1}|\bm{h}m_{\bm{b}_1})$ for brevity) are shown in Tables \ref{tab:cpdF} and \ref{tab:cpdb}, respectively.
	
	\begin{table}[ht]
		\centering
		% table caption is above the table
		\caption{A table showing the entries of the conditional probability distribution $P(F_{\bm{T}}|h\bm{b}_{1},h\bm{b}_{2},h\bm{b}_{3},h\bm{b}_{4})$ where $h\bm{b}=0$ and $h\bm{b}=1$ denote a bay being intact and failed, respectively, and $F_{\bm{T}}=0$ and $F_{\bm{T}}=1$ denote the truss being operational and failed, respectively.}
		\label{tab:cpdF}       % Give a unique label
		% For LaTeX tables use
		\begin{tabular}{cccc|cc}
			\hline\noalign{\smallskip}
			$h\bm{b}_{1}$ & $h\bm{b}_{2}$ & $h\bm{b}_{3}$ & $h\bm{b}_{4}$ & $P(F_{\bm{T}} = 1|\bm{hb})$ & $P(F_{\bm{T}} = 0|\bm{hb})$ \\
			\noalign{\smallskip}\hline\noalign{\smallskip}
			$0$ & $0$ & $0$ & $0$ & $0$ & $1$ \\
			$1$ & $0$ & $0$ & $0$ & $1$ & $0$ \\
			$0$ & $1$ & $0$ & $0$ & $1$ & $0$ \\
			$1$ & $1$ & $0$ & $0$ & $1$ & $0$ \\
			$0$ & $0$ & $1$ & $0$ & $1$ & $0$ \\
			$\vdots$ & $\vdots$ & $\vdots$ & $\vdots$ & $\vdots$ & $\vdots$ \\
			$0$ & $1$ & $1$ & $1$ & $1$ & $0$ \\
			$1$ & $1$ & $1$ & $1$ & $1$ & $0$ \\
			\noalign{\smallskip}\hline
		\end{tabular}
	\end{table}
	
	\begin{table}[ht]
		\centering
		% table caption is above the table
		\caption{A table showing the entries of the conditional probability distribution $P(h\bm{b}_{1}|hm_{9},hm_{13})$ where $hm=0$ and $hm=1$ denote a member being intact and failed, respectively, and $h\bm{b}=0$ and $h\bm{b}=1$ denote a bay being intact and failed, respectively.}
		\label{tab:cpdb}       % Give a unique label
		% For LaTeX tables use
		\begin{tabular}{cc|cc}
			\hline\noalign{\smallskip}
			$hm_{9}$ & $hm_{13}$ & $P(h\bm{b}_{1}=1|\bm{h}m_{\bm{b}_1})$ & $P(h\bm{b}_{1}=0|\bm{h}m_{\bm{b}_1})$\\
			\noalign{\smallskip}\hline\noalign{\smallskip}
			$0$ & $0$ & $0$ & $1$ \\
			$1$ & $0$ & $0$ & $1$ \\
			$0$ & $1$ & $0$ & $1$ \\
			$1$ & $1$ & $1$ & $0$ \\
			\noalign{\smallskip}\hline
		\end{tabular}
	\end{table}
	
	\subsection{Health-state Transition Modelling}\label{sec:HsTM}
	
	The purpose of health-state transition modelling is to develop the conditional probability distribution $P(\bm{H}_{t+1}|\bm{H}_{t},d)$ that predicts the future health state of the truss forward in time, given the current health state and the decided action.
	
	For the purpose of this demonstration, it was decided that the `perform maintenance' action simply returns the structure to its undamaged state, i.e. with no cross-members failed, with probability $1$, independent of $\bm{H}_{t}$.
	
	With regard to the `do nothing' action, it was first assumed that the truss would not spontaneously transition from a more advanced damaged state to a lesser one, that is to say, cross-members would not self-repair in the absence of intervening maintenance, or, without maintenance the health-state of the structure monotonically degrades as a function of time.
	
	The assumed loading range $[0,w_{max}]$ was discretised into 100 evenly-spaced increments; combined with the 8 possible load locations, this resulted in 800 unique considered load cases $L_c$.
	
	For a given load case $L_c$, a transition in health state was defined as $\bm{H}_{t+1} = \bm{H}_{t} + \delta\bm{H}_{t \rightarrow t+1}$ where $\delta\bm{H}_{t \rightarrow t+1}$ is an 8-bit binary vector with $i^{th}$ entry equal to 1 if the yield stress of aluminium (300 MPa) is exceeded in member $m_{i+8}$ when the truss is simulated in health state $H_{t}$ subject to load case $L_c$, and equal to 0 otherwise. The conditional probability of transitioning from $\bm{H}_{t} = H_{t}$ to $\bm{H}_{t+1} = H_{t+1}$, $P(\bm{H}_{t+1} = H_{t+1} | \bm{H}_{t} = H_{t}, L_c)$, was assigned unity if $\delta\bm{H}_{t \rightarrow t+1}=\bm{H}_{t+1}-\bm{H}_{t}$ for $\bm{H}_{t} = H_{t}$ and $\bm{H}_{t+1} = H_{t+1}$ for load case $L_c$, and assigned zero otherwise. The full transition matrix $P(\bm{H}_{t+1}|\bm{H}_{t})$ was then populated where the entry $P(\bm{H}_{t+1}=H_{t+1}|\bm{H}_{t}=H_{t})$ is given by,
	
	\begin{equation}\label{eq:TransitionEq}
	\begin{aligned}
	P(\bm{H}_{t+1}=H_{t+1}|\bm{H}_{t}=H_{t}) = \frac{\sum_{L_c=1}^{N_{L_c}}P(\bm{H}_{t+1} = H_{t+1} | \bm{H}_{t} = H_{t}, L_c)}{N_{L_c}}
	\end{aligned}
	\end{equation}
	
	\noindent where $N_{L_c}$ is the total number of load cases considered and $N_{L_c} = 800$.
	
	For illustrative purposes, the maximum load $w_{max}$ was determined by asserting $P(\bm{H}_{t+1} \ne 0 | \bm{H}_{t} = 0) = 0.005$ and the value of $w_{max}$ that satisfied the condition was found to be approximately 6900 kg. This is obviously somewhat arbitrary, and in practice the maximum load for a structure may be estimated during the operational evaluation stage.
	
\begin{figure}
		\centering
		\includegraphics[scale=0.35]{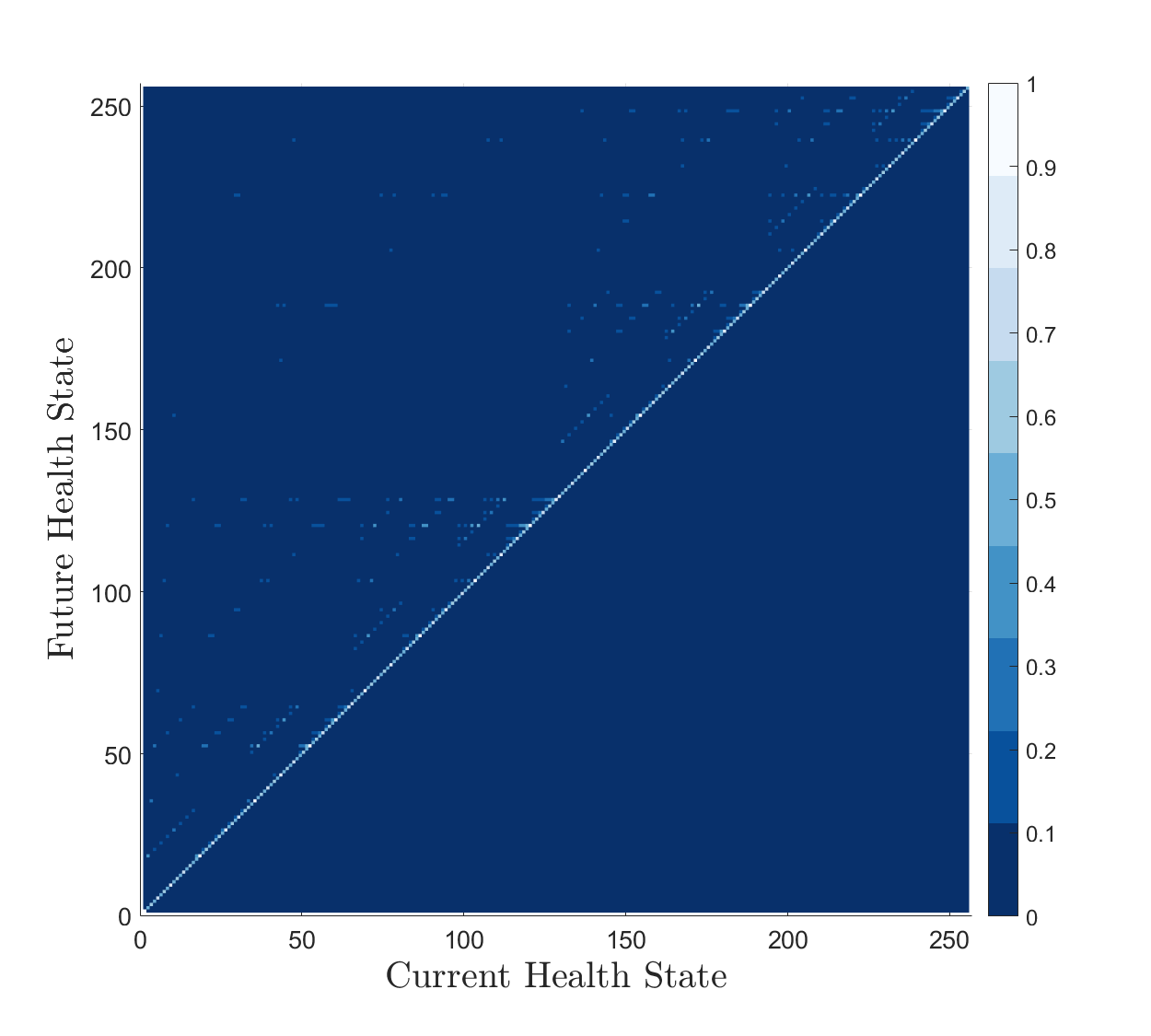}
		\caption{A heatmap showing the conditional probability distribution transition matrix $P(\bm{H}_{t+1}|\bm{H}_{t})$.}
		\label{fig:TransitionMat}
	\end{figure}

	The transition model developed provides a means of forecasting future health states of the truss, a heatmap representation of the transition matrix is shown in Figure \ref{fig:TransitionMat}.
	
	\subsection{Statistical Classifier Development and Training}
	
	The purpose of the statistical classifier is to obtain a probability distribution over the current health state given a set of observed features, i.e. $P(\bm{H}_{t}|\bm{\nu})$. In the current case study, the damage indicative features are the strains measured from the horizontal and vertical members of the truss. The classifier selected for the current case study was comprised of two components; a detector and a localiser, corresponding to the first two stages of Rytter's hierarchy. Whilst generative models may better reflect the causality of the problem at hand, where it is assumed that the features are generated as a result of the latent state and $P(\bm{H}_{t}|\bm{\nu})$ may be computed via Bayes' Theorem, discriminative classifiers that directly learn a mapping from the feature space to the label space are also applicable in the risk-based decision framework.

\begin{figure}
		\centering
		\includegraphics[scale=0.35]{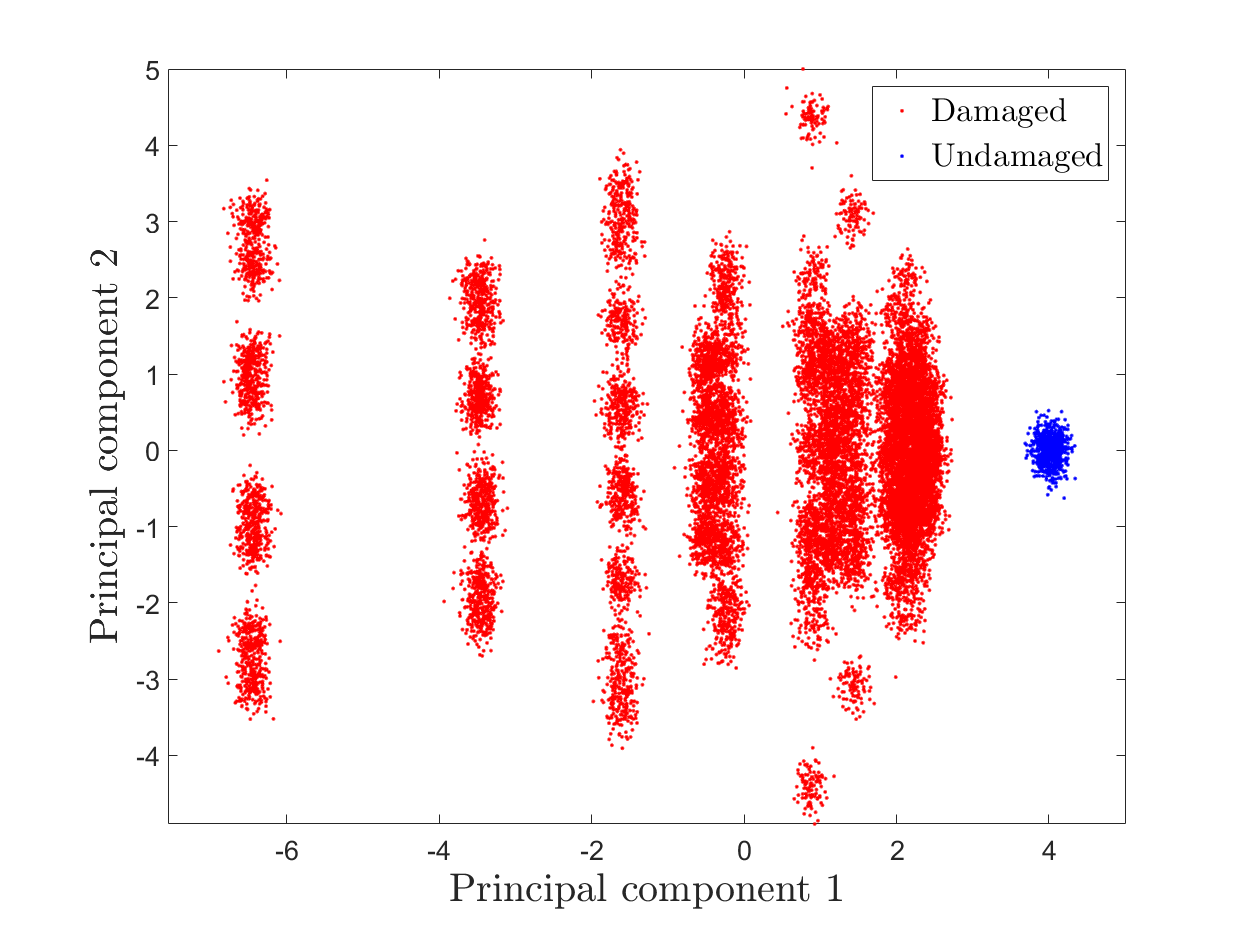}
		\caption{A comparison of the distributions of the first two principal components of the strain data obtained from the finite element model of the undamaged and damaged truss.}
		\label{fig:GMMClusterFig}
	\end{figure}

	A Gaussian novelty detector was implemented to determine the probability that the structure is currently in its undamaged state $P(\bm{H}_t=0|\bm{\nu})$, and, as its complement, the probability that the structure is currently in a damaged state $P(\bm{H}_t\ne0|\bm{\nu})$. The first two principal components of the simulated strain data for the undamaged structure are compared to those from the damaged structure in Figure \ref{fig:GMMClusterFig}. Inspection of Figure \ref{fig:GMMClusterFig} reveals that it is possible to discriminate between the undamaged and damaged finite element simulation data using only the first principal component of the strains hence the novelty detector uses the first principal component as the discriminative feature. As such this principal component projection was learned from the training data and the detector was formed by computing the mean $\mu$ and standard deviation $\sigma$ of the univariate distribution of the first principal component. If the first principal component of an incoming set of strains were to lie within the range $\mu \pm 3\sigma$, it was asserted that the observed strains came from the structure in its undamaged condition with confidence 0.997, i.e. $P(\bm{H}_t=0|\bm{\nu}) = 0.997$. If the first principal component lay outside the  $\pm3\sigma$ confidence interval, $P(\bm{H}_t=0|\bm{\nu})$ was given by the probability mass in the tail of the Gaussian probability density function parametrised by $\mu$ and $\sigma$.
	
	The function of the localiser component of the statistical classifier is to distribute $P(\bm{H}_t\ne0|\bm{\nu})$ over the remaining 255 health states corresponding to the various combinations of cross-member failures. For consistency with \cite{Worden1993}, the classifier selected for this purpose was an artificial neural network (ANN) with an input node for each of the twelve strain measurements, an output node for each cross-member, and three hidden layers with twelve, twelve and eight nodes, respectively. The activation function used was the hyperbolic tangent function. As an identical classifier to that used in \cite{Worden1993} was implemented, only health states 
	$\bm{H}_t= \big\{1,2,4,8,16,32,64,128\big\}$ are considered by the localiser in this case study.
	
	In accordance with \cite{Worden1993}, a training dataset was constructed from the finite element data for the loads of 10, 20 and 30 kg and a validation dataset from the finite element data for the loads of 5, 15 and 25 kg. In both instances, 100 repetitions of the datasets were produced and superimposed with a noise pattern of 1 microstrain RMS. The optimal weights were computed using the scaled conjugate gradient (SCG) back-propagation algorithm \cite{Moller1993} and evaluated and selected based on the classification performance of the network on the validation dataset. Although not inherently probabilistic, a pseudo-probabilistic interpretation for the activations of the output nodes was acquired through the use of a softmax function.

	\subsection{Overall Probabilistic Graphical Model}
	
\begin{figure*}[h!]
		\centering
		\begin{tikzpicture}[x=1.7cm,y=1.8cm]
		
		% Nodes for plate GM
		\node[det] (uf1) {$U_{F_{t+0}}$} ;
		\node[latent,below=1cm of uf1] (failure) {$F_{\bm{T}}$} ;
		\node[const, below=1cm of failure] (temp) {$ $} ;
		\node[latent, left=0.1cm of temp] (bay2) {$h\bm{b}_{2}$} ;
		\node[latent, left=1cm of temp] (bay1) {$h\bm{b}_{1}$} ;
		\node[latent, right=0.1cm of temp] (bay3) {$h\bm{b}_{3}$};
		\node[latent, right=1cm of temp] (bay4) {$h\bm{b}_{4}$} ;
		\node[const, below=1.5cm of temp] (temp2) {$ $} ;
		\node[latent, left=0.1cm of temp2] (mem6) {$hm_{14}$} ;
		\node[latent, left=1.5cm of temp2] (mem2) {$hm_{10}$} ;
		\node[latent, left=3cm of temp2] (mem5) {$hm_{13}$} ;
		\node[latent, left=4.5cm of temp2] (mem1) {$hm_{9}$} ;
		\node[latent, right=0.1cm of temp2] (mem3) {$hm_{11}$} ;
		\node[latent, right=1.5cm of temp2] (mem7) {$hm_{15}$} ;
		\node[latent, right=3cm of temp2] (mem4) {$hm_{12}$} ;
		\node[latent, right=4.5cm of temp2] (mem8) {$hm_{16}$} ;
		\node[det, right=4.6cm of uf1] (uf2) {$U_{F_{t+1}}$} ;
		\node[rectangle,draw=black,dashed,minimum width=0.7cm,minimum height=1cm,below=1cm of uf2] (ft2) {$F^\prime_{t+1}$} ;
		\node[latent, below=2cm of temp2] (x1) {$\bm{H}_{t+0}$} ;
		\node[latent, right=5cm of x1] (x2) {$\bm{H}_{t+1}$} ;
		\node[obs, below=1cm of x1] (y1) {$\bm{\nu}_{t+0}$} ;
		\node[rectangle,draw=black,minimum width=0.7cm,minimum height=1cm,below=1cm of y1] (d1) {$d_{t+0}$} ;
		\node[det, below=1cm of d1] (u1) {$U_{d_{t+0}}$} ;
		\node[rectangle,draw=black,minimum width=0.7cm,minimum height=1cm,right=5cm of d1] (d2) {$d_{t+1}$} ;
		\node[latent, right=1.5cm of x2] (x3) {$\bm{H}_{t+2}$} ;
		\node[det, right=1.1cm of uf2] (uf3) {$U_{F_{t+2}}$} ;
		\node[rectangle,draw=black,dashed,minimum width=0.7cm,minimum height=1cm,below=1cm of uf3] (ft3) {$F^\prime_{t+2}$} ;
		\node[det, below=1cm of d2] (u2) {$U_{d_{t+1}}$} ;
		
		\edge {failure} {uf1} ; %
		\edge {ft2} {uf2} ; %
		\edge {x2} {ft2} ; %
		\edge {x1} {x2} ; %
		\edge {x1} {y1} ; %
		\edge {d1} {x2} ; %
		\edge {d1} {u1} ; %
		\edge {bay1} {failure} ; %
		\edge {bay2} {failure} ; %
		\edge {bay3} {failure} ; %
		\edge {bay4} {failure} ; %
		\edge {mem1} {bay1} ; %
		\edge {mem5} {bay1} ; %
		\edge {mem2} {bay2} ; %
		\edge {mem6} {bay2} ; %
		\edge {mem3} {bay3} ; %
		\edge {mem7} {bay3} ; %
		\edge {mem4} {bay4} ; %
		\edge {mem8} {bay4} ; %
		\edge {x1} {mem1} ; %
		\edge {x1} {mem5} ; %
		\edge {x1} {mem2} ; %
		\edge {x1} {mem6} ; %
		\edge {x1} {mem3} ; %
		\edge {x1} {mem7} ; %
		\edge {x1} {mem4} ; %
		\edge {x1} {mem8} ; %
		\edge {y1} {d1} ; %
		\edge {x2} {x3} ; %
		\edge {x3} {ft3} ; %
		\edge {ft3} {uf3} ; %
		\edge {d2} {x3} ; %
		\edge {d1} {d2} ; %
		\edge {d2} {u2} ; %

		\end{tikzpicture}
		\caption{An influence diagram representing the partially observable Markov decision process for determining the utility-optimal maintenance strategy for the cross-members of a four-bay truss given observations of strains made from the horizontal and vertical members. Observed variables are shaded grey. The fault tree failure models for the latter time steps have been represented as the nodes $F^\prime_{t}$ for compactness.}
		\label{fig:OverallPGM1}
	\end{figure*}
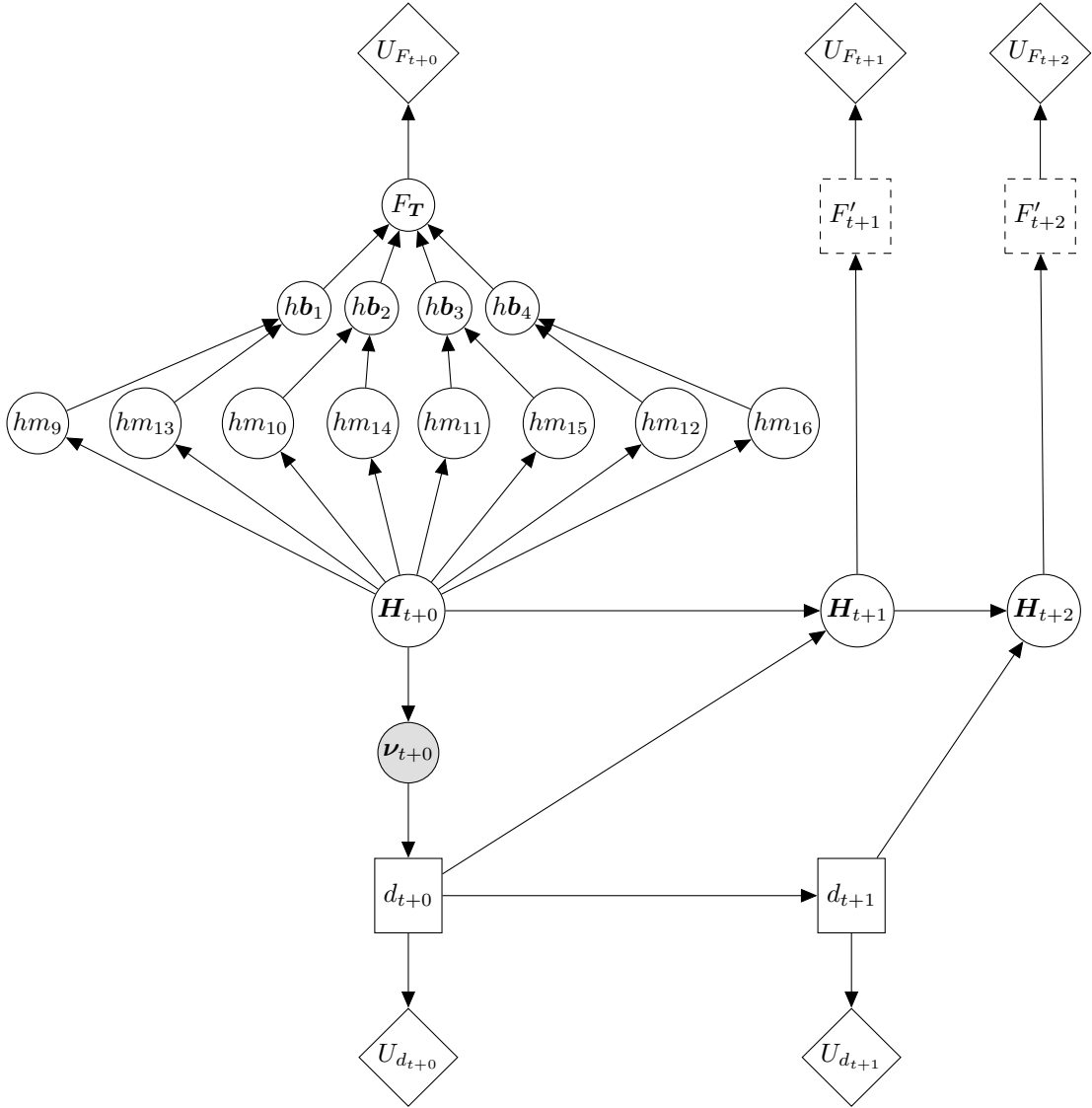

	The failure model, transition models, classifier, decisions and utilities can be combined to form a partially-observable Markov decision process represented by the limited memory influence diagram (LIMID) shown in Figure \ref{fig:OverallPGM1}. Figure \ref{fig:OverallPGM1} shows the decision process for two decisions over three time-slices. The edge connecting $\bm{\nu}_{t+0}$ to $d_{t+0}$ implies that the features are observed prior to the decision being made. Similarly, the edge connecting $d_{t+0}$ to $d_{t+1}$ implies $d_{t+0}$ is decided before $d_{t+1}$.

	It should be noted that the model shown in Figure \ref{fig:OverallPGM1} assumes a generative model for the features $\bm{\nu}$; for discriminative classifiers the direction of the conditioning edge connecting nodes $\bm{H}_{t+0}$ and $\bm{\nu}_{t+0}$ would be reversed.

	\section{Results}
	
	\subsection{Novelty Detector}
	
\begin{figure}
		\centering
		\includegraphics[scale=0.6]{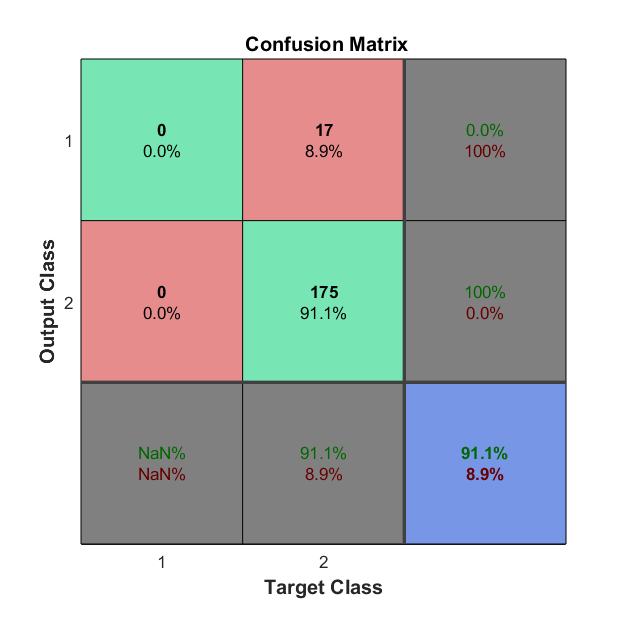}
		\caption{A confusion matrix detailing the classification performance of the novelty detector on the experimental data. }
		\label{fig:detectorConfusion}
	\end{figure}

	When applied to the experimental strain data from the damaged truss, the univariate Gaussian novelty detector was able to correctly identify 175 out of the 192 observations as novel with respect to the simulated undamaged data, thereby yielding an overall accuracy of 91.1\% and shown in the confusion matrix in Figure \ref{fig:detectorConfusion}.
	
\begin{figure}
		\centering
		\includegraphics[scale=0.35]{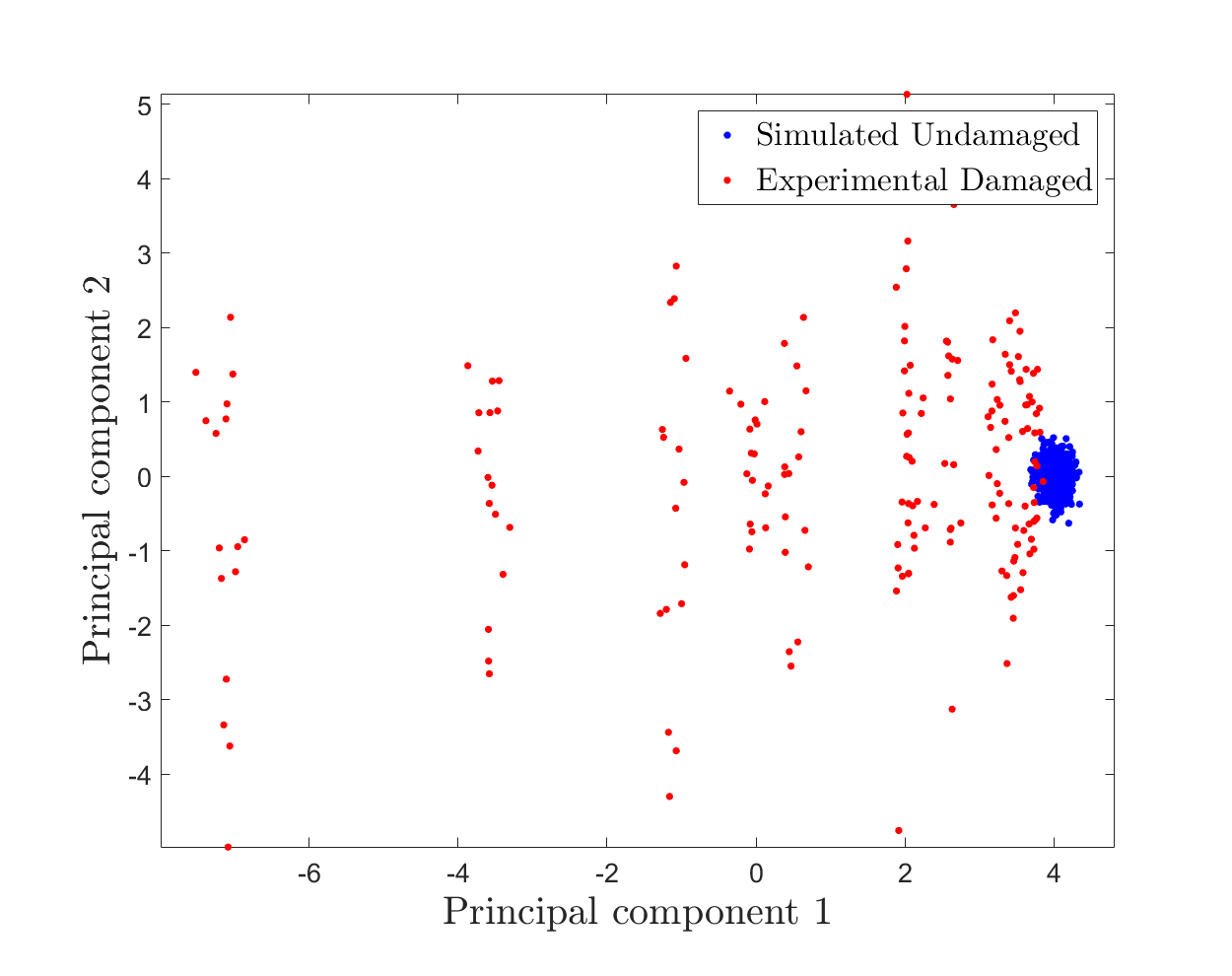}
		\caption{A comparison of the distributions of the first two principal components of the strain data obtained from the finite element model of the undamaged truss and the strain data from the experiment performed on the damaged truss mapped through the projection learned from the simulated training data.}
		\label{fig:detectorClusters}
	\end{figure}

	The 8.9\% misclassification can be elucidated by examining the distribution of the experimental data as projected through the principal component mapping learned from the finite element simulation data, shown in Figure \ref{fig:detectorClusters}. Separability between the projected damaged and undamaged datasets is absent where values of the first principal component are approximately 3.9.
	
	\subsection{Localiser}
	
\begin{figure}
		\centering
		\includegraphics[scale=0.6]{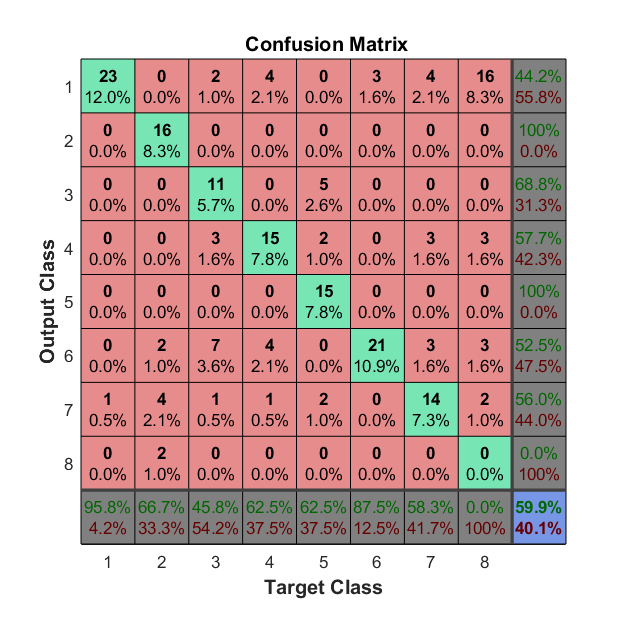}
		\caption{A confusion matrix detailing the classification performance of the artificial neural network localiser on the experimental data.}
		\label{fig:localiserConfusion}
	\end{figure}

	When applied to the experimental strain data, the neural network localiser has an overall classification accuracy of 59.9\%. The full confusion matrix is shown in Figure \ref{fig:localiserConfusion}. Whilst 60\% classification accuracy may be considered low, for a 8 class problem the neural network provides a significant improvement over simply guessing which would yield an accuracy of 12.5\%. The imperfect classifier has been deliberately chosen here as a possible source of uncertainty.
	
	The misclassification error of 40.1\% can be explained by considering the physics of the problem at hand. The selected damage sensitive features were the strains in the horizontal and vertical members of the truss; however, significant changes in the strains are only expected of members in the load path between the end fixture and applied mass. Therefore, the features are largely insensitive to damage when the mass is closer to the fixture than the damaged cross-member and any differences can be attributed to the strains induced by the preload. In 72 of the 192 (37.5\%) cases, the damaged cross-member is not in the load path.
	
	%\begin{figure}
	%	\includegraphics[scale=0.27]{DamageDataScatter1.png}
	%	\caption{A plot comparing the distributions of the first two principal components of the strain data obtained from the experiment performed on the damaged truss. Data originating from each damage condition are differentiated via colour with the legend number corresponding the failed cross-member's number.}
	%	\label{fig:localiserClusters1}
	%\end{figure}
	%
	%\begin{figure}
	%	\includegraphics[scale=0.27]{DamageDataScatter2.png}
	%	\caption{A plot comparing the distributions of the second and third principal components of the strain data obtained from the experiment performed on the damaged truss. Data originating from each damage condition are differentiated via colour with the legend number corresponding the failed cross-member's number.}
	%	\label{fig:localiserClusters2}
	%\end{figure}

	\subsection{Decision Process Results}
	
	The decision algorithm was tested on a dataset comprised of the experimental strains for the truss in its damaged conditions and, due to the lack of experimental data, finite element simulation data for the truss in its undamaged condition. Equal proportions of the undamaged and damaged data were used; 192 sets of strains from each.
	
	The decision algorithm used for testing was similar to that shown in Figure \ref{fig:OverallPGM1}, except using the discriminative pseudo-probabilistic ANN classifier rather than a generative model. The graphical was modelled in MATLAB using the Bayes Net Toolbox \cite{Murphy2001} and solved using the junction tree algorithm for influence diagrams described in \cite{Jensen1994}. Utilities were as shown in Tables \ref{tab:UF} and \ref{tab:Ud}. Three failure events in consecutive time-steps were considered in conjunction with two `do nothing'/`perform maintenance' decisions in the first two time-steps with a single observation made during the first time-step.
	
	\begin{figure}
		\centering
		\includegraphics[scale=0.6]{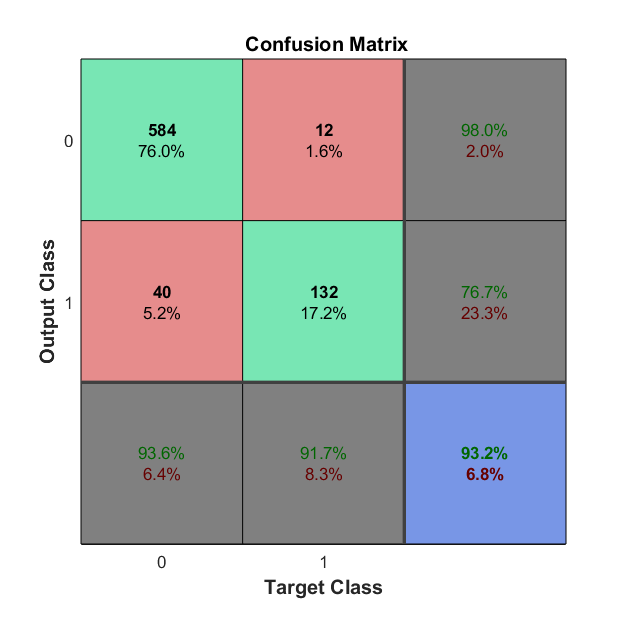}
		\caption{A confusion matrix detailing the `accuracy' of the decision algorithm for both decisions in the three time-slice problem for $U(F_{\bm{T}}=1)=-285$.}
		\label{fig:AllDecisionConfusion}
	\end{figure}
	
\begin{figure}
		\centering
		\includegraphics[scale=0.6]{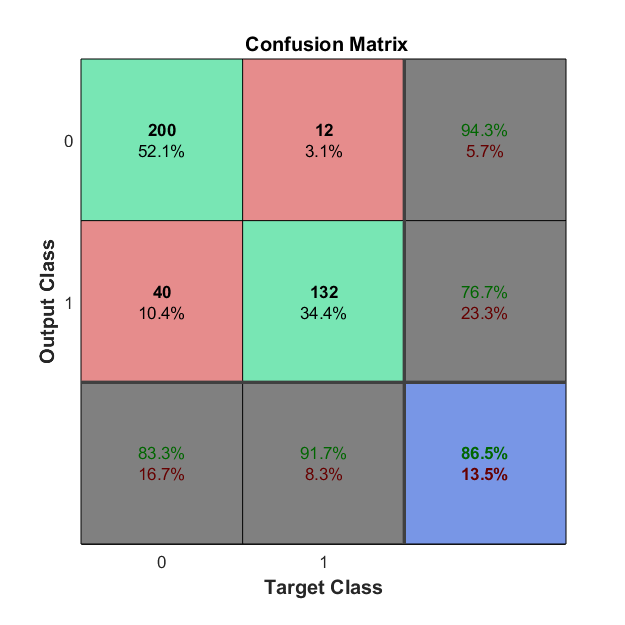}
		\caption{A confusion matrix detailing the `accuracy' of the decision algorithm for the first decision in the three time-slice problem for $U(F_{\bm{T}}=1)=-285$.}
		\label{fig:FirstDecisionConfusion}
	\end{figure}

\begin{figure}
		\centering
		\includegraphics[scale=0.6]{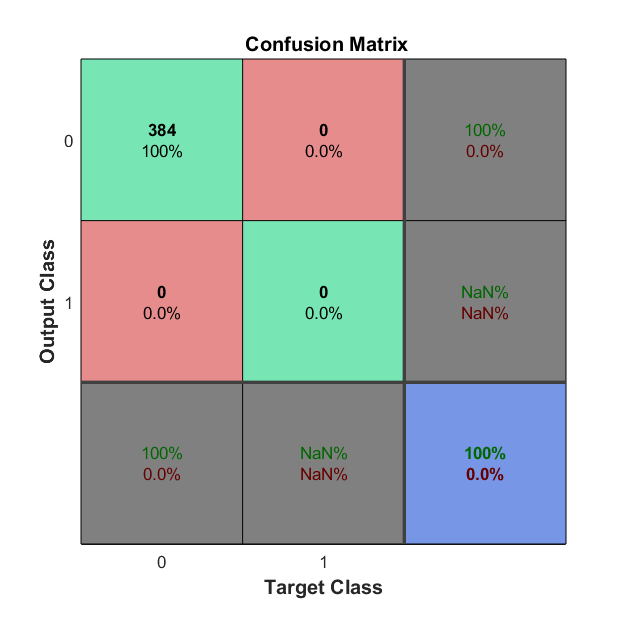}
		\caption{A confusion matrix detailing the `accuracy' of the decision algorithm for the second decision in the three time-slice problem for $U(F_{\bm{T}}=1)=-285$.}
		\label{fig:SecondDecisionConfusion}
	\end{figure}

	The performance of the decision process was evaluated with a metric similar to that of a classifier's overall accuracy. Whereas a classifiers accuracy is a comparison between the predicted outputs and the target outputs, this `decision accuracy' is a comparison between the decided actions given the outputs of the classifier used to the optimal decided actions given perfect information of the health state, i.e. the utility-optimal decisions when the targets of the classifier are provided in place of $P(\bm{H}|\bm{\nu})$. The target and output classes `0' and `1' for Figures \ref{fig:AllDecisionConfusion}, \ref{fig:FirstDecisionConfusion} and \ref{fig:SecondDecisionConfusion} correspond to the `do nothing' and `perform maintenance' actions, respectively.
	
	Figure \ref{fig:AllDecisionConfusion} shows the performance of the decision algorithm across all 786 decisions associated with the test dataset. It can be seen that an overall `accuracy' of 93.2\% was achieved meaning that the optimal decision given perfect information of the health state selected in 716 of the cases when the statistical classifier was used to infer the health state. In 40 of the cases, the `perform maintenance' action was selected unnecessarily, this is a result of the uncertainty in the health state triggering a more, conservative action to be taken; this form of error is akin to a `false positive' or type I error. In 12 of the cases the `do nothing' action was deemed to be optimal whereas, had perfect information of the structures health state been available, the optimal decision would, in fact, have been `perform maintenance'. This form of error is akin to a `false negative' or type II error.
	
	The severity/significance of type I and type II errors is dependent on the context of the SHM system. For example, for an offshore wind structure, erroneously sending inspection/maintenance engineers has a higher cost relative to the failure event, whereas for a bridge the cost of inspection/maintenance is relatively lower with respect to the cost of failure. In this risk-based decision-framework, the costs are explicitly modelled and may be used to inform a preferential selection of classifier with regard to type I and type II errors.

	Figures \ref{fig:FirstDecisionConfusion} and \ref{fig:SecondDecisionConfusion} provide a breakdown of the `decision accuracy' shown in Figure \ref{fig:AllDecisionConfusion} for the first and second decisions, respectively. Figure \ref{fig:FirstDecisionConfusion} shows that, for the first decision, the algorithm was able to select the correct action in 332 of the 384 cases, yielding an overall `decision accuracy' of 86.5\%. Additionally, it can be seen that in 40 cases the `perform maintenance' action was selected incorrectly, and in 12 cases the `do nothing' action was selected incorrectly. Comparing Figures \ref{fig:AllDecisionConfusion} and \ref{fig:FirstDecisionConfusion} reveals that all type I and type II errors occur during the first decision. Logically, Figure \ref{fig:SecondDecisionConfusion} shows that the algorithm has a `decision accuracy' of 100\% with respect to the second decision. Moreover, Figure \ref{fig:SecondDecisionConfusion} shows that all optimal decisions during the second time slice are `do nothing'. This result can be explained by considering two possibilities. During the first time step, if the algorithm has decided that maintenance is warranted, then, under the assumed transition model $P(\bm{H}_{t+1}|\bm{H}_{t+0},d=1)$, the structure is guaranteed to be in its undamaged health state in the second time step in which case further maintenance is unwarranted given $P(\bm{H}_{t+2}|\bm{H}_{t+1},d=0)$ and $U(F_{\bm{T}}=1)=-285$. Alternatively, if, during the first time step, the algorithm deems `do nothing' to be the optimal decision, then the degradation forecast over two time steps is not sufficient to warrant a `perform maintenance' action in the second time step; however, this decision, of course, is dependent on the utilities specified within the model.
	
	\begin{figure}
		\centering
		\includegraphics[scale=0.35]{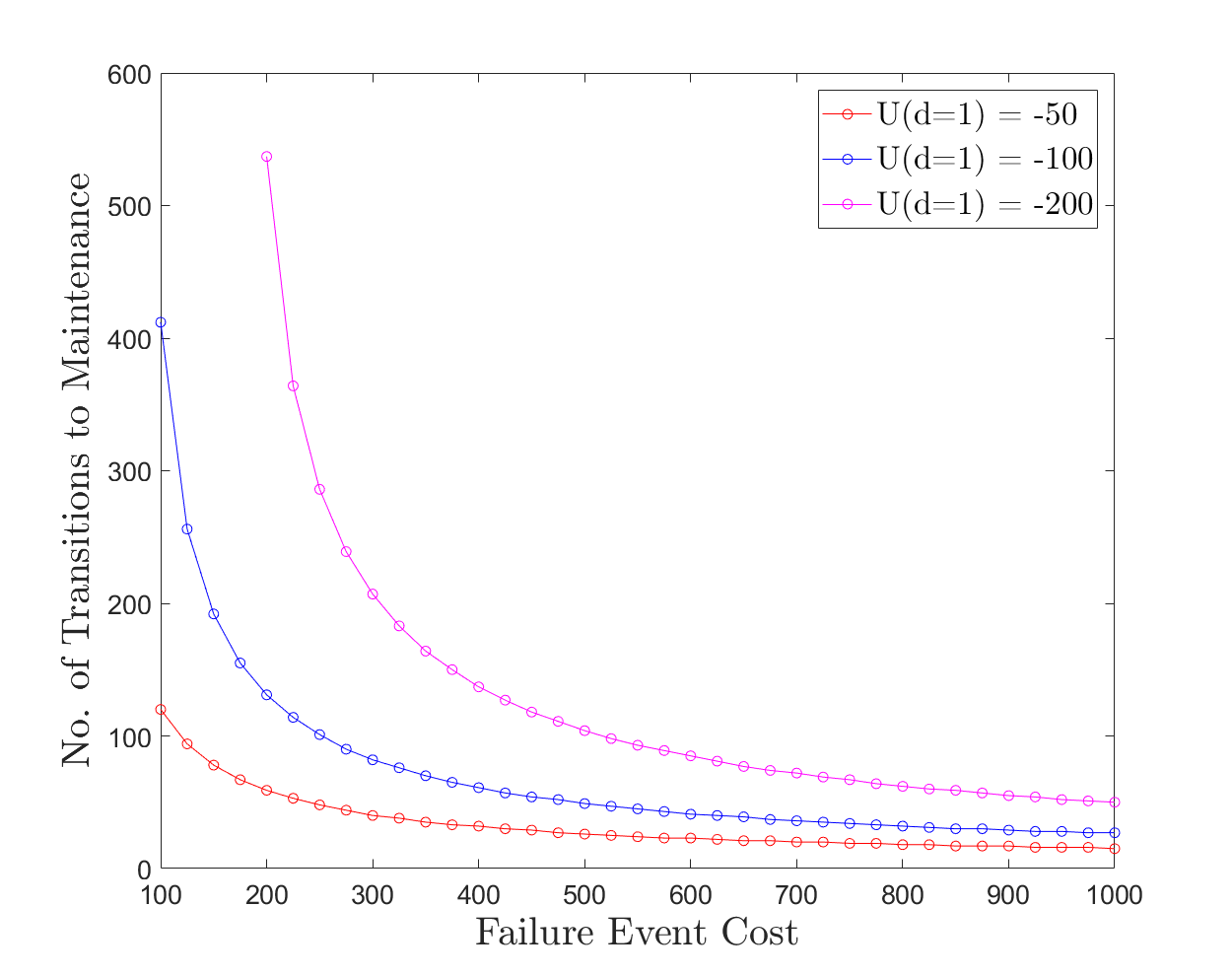}
		\caption{The variation in the number of state transitions the undamaged structure will go through before maintenance is decided as a function of the failure event cost and maintenance cost. Failure event cost is defined as $-U(F_{\bm{T}}=1)$.}
		\label{fig:TimeVCost}
	\end{figure}
	
	Figure \ref{fig:TimeVCost} shows the number of state transitions an initial undamaged state may go through before a `perform maintenance' action is decided when the cost of the failure event and cost of maintenance are varied and all other utilities remain as specified in Tables \ref{tab:UF} and \ref{tab:Ud}. It can be seen that the number of transitions decays exponentially as the cost of failure increases. This is in accordance with the intuitive understanding that as the cost of failure tends to infinity, the number of transitions before maintenance should be decided asymptotically approaches zero. It can also be seen from Figure \ref{fig:TimeVCost} that for a given failure event cost the time until maintenance decreases with cost of maintenance. It should also be noted that, logically, if the cost of maintenance exceeds the cost of failure, the structure will be allowed to operate until failure.
	
	\begin{figure}
		\centering
		\includegraphics[scale=0.35]{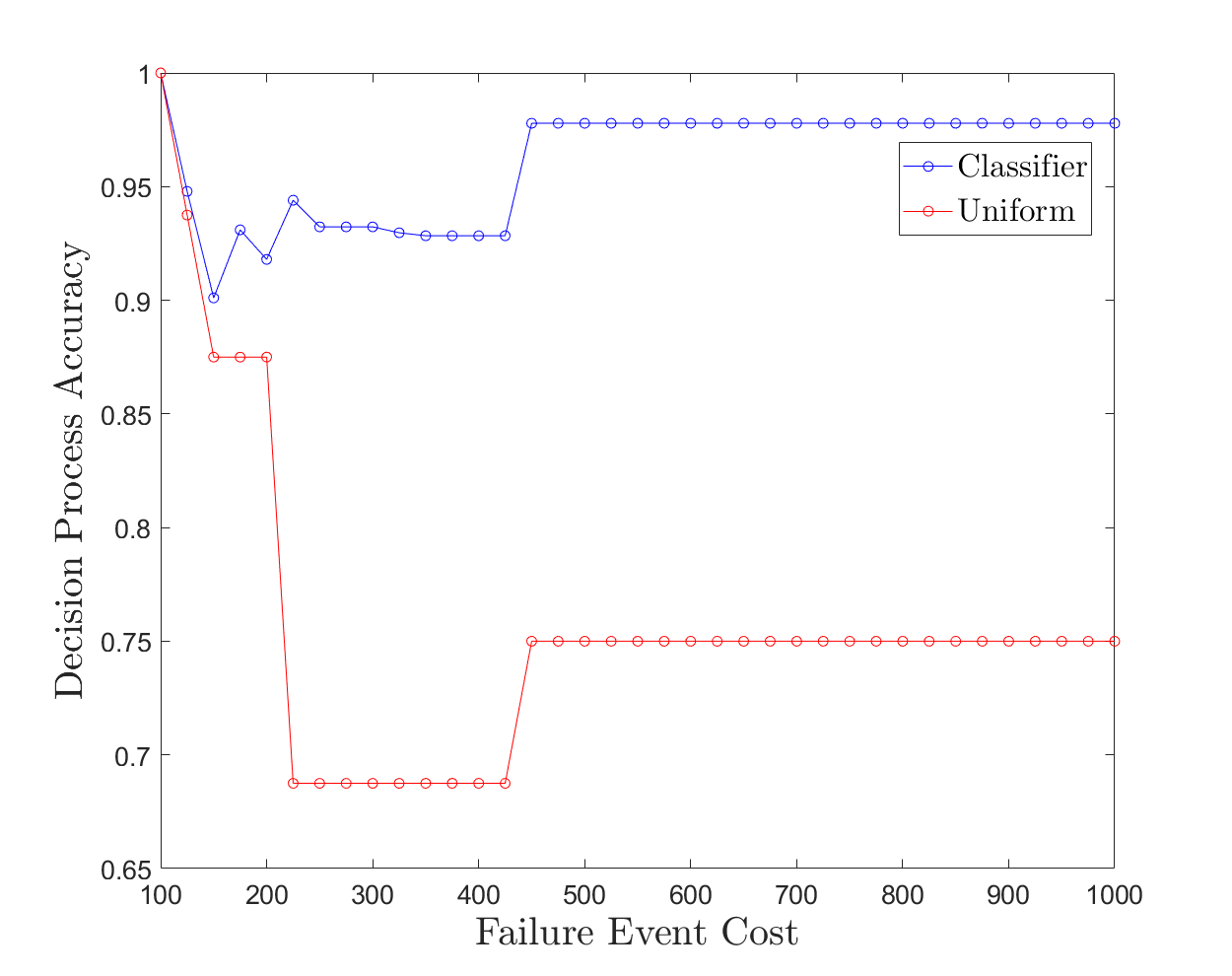}
		\caption{A comparison of the `accuracy' of the decision process as a function of the cost of the failure event when the health state is inferred using the statistical classifier and a uniform distribution over health states is assumed. Failure event cost is defined as $-U(F_{\bm{T}}=1)$.}
		\label{fig:AccuracyVsCost}
	\end{figure}
	
	To investigate the influence of the cost of failure upon the overall `accuracy' of the decision algorithm, the decision process used to produce Figure \ref{fig:AllDecisionConfusion} was repeated for varying $U(F_{\bm{T}}=1)$. It should be noted that the utility of `perform maintenance' action was fixed at $U(d=1)=-100$. In addition to varying $U(F_{\bm{T}}=1)$, the decision algorithm was executed assuming a uniform distribution over the health states targeted by the classifier rather than the distribution as predicted by the classifier. Figure \ref{fig:AccuracyVsCost} shows how the `decision accuracy' of each algorithm varies with the cost of the failure event.
	
	Figure \ref{fig:AccuracyVsCost} shows that the decision algorithms using the classifier and the uniform distribution assumption all have perfect `accuracy' when $U(F_{\bm{T}}=1)=-100$. This result can be attributed to the fact that when the cost of maintenance is less than or equal to the cost of failure, the optimal decision is always `perform maintenance', therefore, the algorithms all select this action independent of the information available regarding the health state of the structure.
	
	It can be seen from Figure \ref{fig:AccuracyVsCost} that, with the exception of the somewhat trivial case of $U(F_{\bm{T}}=1)=-100$, the decision algorithm utilising the classifier to incorporate probabilistic information about the health state of the structure into the decision consistently performs better than assuming a uniform distribution over health states. This is to be expected as the use of the classifier allows for the identification of the undamaged condition and differing damage states. On the other hand, as assuming a uniform distribution is entirely ignorant of the health state the selected decision will be invariant for a given failure event cost. Therefore, even when the structure is undamaged, assuming a uniform distribution will result in a maintenance action to be selected whilst the failure event cost is sufficiently high.
	
	The `accuracies' of all three algorithms follow roughly the same trend for the range of failure costs shown in Figure \ref{fig:AccuracyVsCost}. For $U(F_{\bm{T}}=1)\le-450$  the accuracy is constant. This is likely because, due to the high cost of failure, given the perfect knowledge of the health state, the optimal decision for all damage cases other than undamaged are `perform maintenance'. For the algorithm utilising the classifier, misclassification of damage location does not influence the decided action in this cost range. As previously mentioned, the uniform assumption is ignorant of the health state and the decided actions are invariant in the range $U(F_{\bm{T}}=1)\le-450$.
	
	In Figure \ref{fig:AccuracyVsCost}, a lower `decision accuracy' is seen in the failure cost range $-450<U(F_{\bm{T}}=1)\le-225$. This is expected to be because some of the optimal decisions given perfect information of the health for certain damage locations to be `do nothing'. The uncertainty in health state reflected in the classifier and uniform distribution algorithms cause the `perform maintenance' action to be deemed appropriate.
	
	In the failure cost range $U(F_{\bm{T}}=1)\ge-200$ an increase in accuracy is seen for the algorithm assuming a uniform distribution over the health states. This is a result of the health state invariant decision being `do nothing'; in this failure cost range this assumption is able to correctly decide actions for the undamaged cases and the less severe damage locations. The `decision accuracy' of the algorithm employing the classifier fluctuates in this range, this may be due to the decided actions being sensitive to the distribution of uncertainties over the health states.
	
	\section{Discussion}
	
	%Discussion points:
	%\begin{itemize}
	%	\item Utility gains can still be made regardless of the classifiers overall accuracy provided uncertainty is dealt with suitably... Desirable to move towards true probabilistic classifiers, be they discriminative (RVMs) or generative (GMMs).
	%	\item Decisions are able to be made regarding local components/substructures whilst taking into account the effect on failure modes of the global structure thanks to fault trees.
	%	\item Care must be taken in assigning costs and utilities.
	%	\item Transition model is very important.
	%	\item Validation is hard for one-off structures due to interventions. May be easier for operating populations of homogeneous structures. Validation for decisions and for models within algorithm.
	%	\item bring back to positive
	%	
	%\end{itemize}
	
	The framework described and demonstrated in the current paper provides an approach to risk-based decision-making in the context of SHM. Decision-making is facilitated through the inclusion of aspects of PRA such as fault tree modelling and risk, thereby allowing for the comparison of actions and the identification of a strategy that maximises expected utility.
	
	The PRA paradigm currently practiced in industries such as aerospace and nuclear provides a basis for the formalisation of the operational evaluation procedure. Organising the information specifying the structure and monitoring system in a database will assist with ensuring all the necessary information required for subsequent stages is acquired and it will also provide a structured method for the retrieval of applicable information at each stage.
	
	The fault tree development process of PRA provides the key novelty of this risk-based approach to SHM. Firstly, it facilitates the definition of key failure modes of interest and provides a structured method for identifying pertinent components whose health states should be targeted by a statistical classifier. The size of decision space for any given structure in the context of SHM is vast and an intimidating problem to begin addressing. By targeting selected failure modes of interest for a structure and modelling them as fault trees, the scope of the decision-maker may be limited, thereby making the problem more approachable; additional failure modes may subsequently be incorporated as an SHM system is further developed/expanded. Mapping the fault trees into Bayesian networks enables the framework to retain information regarding the uncertainties in the health states thereby allowing robustness in the decision-making. Moreover, intermediate nodes within the Bayesian network representation of a fault tree may be queried, yielding marginal distributions that provide information about the probability of damage within components and/or substructures. This information may be utilised to guide inspection and maintenance engineers to specific locations, potentially saving time and reducing the cost of the actions, particularly for larger structures.
	
	Whilst the framework presented addresses some of the problems surrounding the SHM decision process, there remain a number of challenges. One challenge, that has been widely acknowledged in the SHM community, is that data from the damage states of interest for a structure are seldom available prior to the implementation of an SHM system. This poses an issue in the development of the classifiers on which the decision process is highly dependent and a choice must be made regarding the approach to the statistical modelling.  One option is to take a model-driven approach \cite{Barthorpe2011} that utilises outputs from physics-based models of the structure in its damage states of interest to learn a classifier in a supervised manner pre-implementation of the SHM system. Subsequently, the classifier can be continuously updated and validated with data obtained during the monitoring campaign. Alternatively, a semi-supervised approach can be taken in which a clustering algorithm is applied to the data acquired throughout the monitoring campaign. Clusters are attributed damage state labels through the incorporation of labelled data into the clustering algorithm \cite{Bull2019}; damage state labels for data points may be obtained through inspection of the structure \cite{Rogers2019}.
	
	The results presented previously show that the performance of a probabilistic risk-based decision algorithm is dependent on the available information regarding the health state of the structure. As demonstrated, gains with regards to utility can be made in the absence of high-accuracy classifiers provided uncertainty is accounted for. This provides motivation for moving towards the use of true probabilistic classifiers in the context of SHM, be they discriminative (such as relevance vector machines (RVMs) \cite{Tipping2001}) or generative (such as GMMs \cite{Bishop2006}).
	
	In addition to being dependent on the statistical classifier used, the optimality of decisions is highly contingent on the appropriateness of the transition model used; if the degradation of the structure is not accurately modelled, erroneous actions may be taken. Facing a similar issue to the statistical modelling process, oftentimes, data describing the transitions between the health states of interest are not held \textit{a priori}. Again, one is faced with the choice of taking a model-driven approach involving the simulation of the degradation, or a data-driven approach that utilises data obtained during the monitoring campaign. The development of data-driven transition models, or validation of model-driven transition models is an awkward problem. Due to the fact that one is performing interventions on the structure during operation, information on transitions between health states is regularly censored, meaning the quantities of data spanning all state transitions of interest required for developing/validating transitions models may never be acquired, and particularly troublesome for one-off/bespoke structures. This problem is left as an open research question.
	
	To ensure the desired performance of the decision algorithm, a vital stage in the risk-based framework is to assign utilities/costs to failure events and actions. Currently, within the literature there is no formal approach to how these values should be elicited, nor is there a consensus on how the risk preferences of an SHM decision-maker should be specified; should an agent be risk averse, risk neutral, or risk seeking? The issue at hand is one of both a technical and ethical nature, and whilst it will not be discussed in further detail in the current paper, it is highlighted to stimulate the conversations required for progress in the area of risk-informed decision-making for SHM.
	
	In summary, a probabilistic risk-based framework for structural health monitoring was presented. Borrowing practices frequently used in probabilistic risk assessment, such as the use of fault trees to  model system failures, the framework facilitates robust decision-making under uncertainty and provides advancements in the utility-optimal operation and maintenance of structures.
	
	\section*{Acknowledgements}
		The authors would like to acknowledge the support of the UK EPSRC via the Programme Grant EP/R006768/1. KW would also like to acknowledge support via the EPSRC Established Career Fellowship EP/R003625/1. The authors would like to thank Dr Mark Bateman of EDF Energy for providing valuable discussions.

	% Authors must disclose all relationships or interests that 
	% could have direct or potential influence or impart bias on 
	% the work: 
	%
	\section*{Conflict of interest}
	
	The authors declare that they have no conflict of interest.

	% BibTeX users please use one of
	%\bibliographystyle{spbasic}      % basic style, author-year citations
	\bibliographystyle{elsarticle-num}
	\bibliography{RiskSHMJournal}
	
\end{document}